\documentclass{emulateapj}


\usepackage{epsfig}             
\usepackage{xspace}
\usepackage{graphicx}
\usepackage{natbib}
\usepackage{bm}
\usepackage{footmisc}
\usepackage{nicefrac}
\usepackage{gensymb}
\usepackage{amsmath}

\begin{document}

\newcommand{\ignore}[1]{}
\newcommand{\nagn}{NGC 5548\xspace}
\newcommand{\perpix}{pixel\ensuremath{^{-1}}\xspace}
\newcommand{\halpha}{H\ensuremath{\alpha}\xspace}
\newcommand{\hbeta}{H\ensuremath{\beta}\xspace}
\newcommand{\hgamma}{H\ensuremath{\gamma}\xspace}
\newcommand{\hdelta}{H\ensuremath{\delta}\xspace}
\newcommand{\hei}{\ion{He}{1}\xspace}
\newcommand{\heii}{\ion{He}{2}\xspace}
\newcommand{\heiiopt}{\ion{He}{2}\xspace$\lambda$4686\xspace}
\newcommand{\oiii}{[\ion{O}{3}]\xspace}
\newcommand{\feii}{\ion{Fe}{2}\xspace}
\newcommand{\siiv}{\ion{Si}{4}\xspace}
\newcommand{\civ}{\ion{C}{4}\xspace}
\newcommand{\lya}{Ly\ensuremath{\alpha}\xspace}
\newcommand{\mgii}{\ion{Mg}{2}\xspace}
\newcommand{\javelin}{\texttt{JAVELIN}\xspace}
\newcommand{\angstrom}{\AA\xspace}
\newcommand{\lam}{\ensuremath{\lambda}}
\newcommand{\lagn}{\ensuremath{L_{\mathrm{AGN}}}\xspace}
\newcommand{\kms}{km s\ensuremath{^{-1}}\xspace}
\newcommand{\ergscm}{erg s$^{-1}$ cm$^{-2}$\xspace}
\newcommand{\nev}{\ensuremath{\sigma^2_{\mathrm{rms}}}\xspace}
\newcommand{\ffiftyone}{\ensuremath{F_{\lambda\mathrm{5100}}}\xspace}
\newcommand{\lfiftyone}{\ensuremath{L_\mathrm{5100}}\xspace}
\newcommand{\lambdalfiftyone}{\ensuremath{\lambda L_\mathrm{5100}}\xspace}
\newcommand{\msun}{\ensuremath{M_{\odot}}\xspace}
\newcommand{\mbh}{\ensuremath{M_{\mathrm{BH}}}\xspace}
\newcommand{\rl}{\ensuremath{R_\mathrm{BLR}-L_\mathrm{AGN}}\xspace}
\newcommand{\rmax}{\ensuremath{r_\mathrm{max}}\xspace}
\newcommand{\Rmax}{\ensuremath{R_\mathrm{max}}\xspace}
\newcommand{\tchar}{\ensuremath{\tau_{\mathrm{char}}}\xspace}
\newcommand{\taupeak}{\ensuremath{\tau_{\mathrm{peak}}}\xspace}
\newcommand{\taucen}{\ensuremath{\tau_{\mathrm{cen}}}\xspace}
\newcommand{\taucenTone}{\ensuremath{\tau_{\mathrm{cen,T1}}}\xspace}
\newcommand{\taucenTtwo}{\ensuremath{\tau_{\mathrm{cen,T2}}}\xspace}
\newcommand{\taujav}{\ensuremath{\tau_{\mathrm{\texttt{JAVELIN}}}}\xspace}
\newcommand{\taujavTone}{\ensuremath{\tau_{\mathrm{\texttt{JAVELIN},T1}}}\xspace}
\newcommand{\taujavTtwo}{\ensuremath{\tau_{\mathrm{\texttt{JAVELIN},T2}}}\xspace}
\newcommand{\taufiftyone}{\ensuremath{\tau_\mathrm{5100}}\xspace}
\newcommand{\tauoptical}{\ensuremath{\tau_\mathrm{\hbeta-opt}}\xspace}
\newcommand{\tauuv}{\ensuremath{\tau_\mathrm{\hbeta-UV}}\xspace}
\newcommand{\tauuvopt}{\ensuremath{\tau_\mathrm{opt-UV}}\xspace}
\newcommand{\fluxunit}{\ensuremath{10^{-15}~\mathrm{erg~s}^{-1}~\mathrm{cm}^{-2}}\xspace}
\newcommand{\fluxunitlambda}{\ensuremath{10^{-15}~\mathrm{erg~s}^{-1}~\mathrm{cm}^{-2}~\mathrm{\angstrom}^{-1}}\xspace}
\newcommand{\msigma}{\ensuremath{M_{\mathrm{BH}}-\sigma_{\star}}\xspace}
\newcommand{\rblr}{\ensuremath{R_{\mathrm{BLR}}}\xspace}
\newcommand{\hst}{\textit{HST}\xspace}
\newcommand{\swift}{\textit{Swift}\xspace}
\newcommand{\vband}{\textit{V}-band\xspace}
\newcommand{\eff}{\ensuremath{\eta_\mathrm{eff}}\xspace}

\title{Space Telescope and Optical Reverberation Mapping Project. V. Optical Spectroscopic Campaign and Emission-Line Analysis for \nagn}
\author{
  L.~Pei\altaffilmark{1,2}, %UCI
  M.~M.~Fausnaugh\altaffilmark{3}, %OSU
  A.~J.~Barth\altaffilmark{1}, %UCI
  B.~M.~Peterson\altaffilmark{3,4,5}, %OSU, CCAP, STScI
  M.~C.~Bentz\altaffilmark{6}, %Georgia
  G.~De~Rosa\altaffilmark{5}, %OSU, CCAP, STScI
  K.~D.~Denney\altaffilmark{3,4,7}, %OSU, CCAP, NSF
  M.~R.~Goad\altaffilmark{8}, %Leicester
  C.~S.~Kochanek\altaffilmark{3,4}, %OSU, CCAP
  K.~T.~Korista\altaffilmark{9}
% ...
}

\altaffiltext{1}{\ignore{UCI}Department of Physics and Astronomy, 4129 Frederick Reines Hall, University of California, Irvine, CA 92697, USA}
\altaffiltext{2}{\ignore{UIUC}Department of Astronomy, University of Illinois at Urbana-Champaign, Urbana, IL 61801, USA}
\altaffiltext{3}{\ignore{OSU}Department of Astronomy, The Ohio State University, 140 W 18th Ave, Columbus, OH 43210, USA}
\altaffiltext{4}{\ignore{CCAPP}Center for Cosmology and AstroParticle Physics, The Ohio State University, 191 West Woodruff Ave, Columbus, OH 43210, USA}
\altaffiltext{5}{\ignore{STScI}Space Telescope Science Institute, 3700 San Martin Drive, Baltimore, MD 21218, USA}
\altaffiltext{6}{\ignore{Georgia}Department of Physics and Astronomy, Georgia State University, 25 Park Place, Suite 605, Atlanta, GA 30303, USA}
\footnotetext[7]{NSF Postdoctoral Research Fellow}
\altaffiltext{8}{\ignore{Leicester}Department of Physics and Astronomy, University of Leicester,  Leicester, LE1 7RH, UK}
\altaffiltext{9}{\ignore{WM}Department of Physics, Western Michigan University, 1120 Everett Tower, Kalamazoo, MI 49008, USA}
\altaffiltext{10}{\ignore{JH}Department of Physics and Astronomy, The Johns Hopkins University, Baltimore, MD 21218, USA}
\altaffiltext{11}{\ignore{CalPoly}Physics Department, California Polytechnic State University, San Luis Obispo, CA 93407, USA}
\altaffiltext{12}{\ignore{Wyoming}Department of Physics and Astronomy, University of Wyoming, 1000 E. University Ave. Laramie, WY 82071, USA}
\altaffiltext{13}{\ignore{Berkeley}Department of Astronomy, University of California, Berkeley, CA 94720-3411, USA}
\altaffiltext{14}{\ignore{galilei}Dipartimento di Fisica e Astronomia ``G. Galilei,'' Universit\`{a} di Padova, Vicolo dell'Osservatorio 3, I-35122 Padova, Italy}
\altaffiltext{15}{\ignore{INAF}INAF-Osservatorio Astronomico di Padova, Vicolo dell'Osservatorio 5 I-35122, Padova, Italy}
\altaffiltext{16}{\ignore{Princeton}Department of Astrophysical Sciences, Princeton University, Princeton, NJ 08544, USA}
\altaffiltext{17}{\ignore{Eberly}Department of Astronomy and Astrophysics, Eberly College of Science, The Pennsylvania State University, 525 Davey Laboratory, University Park, PA 16802, USA}
\altaffiltext{18}{\ignore{IGC}Institute for Gravitation and the Cosmos, The Pennsylvania State University, University Park, PA 16802, USA}
\altaffiltext{19}{\ignore{Dark}Dark Cosmology Centre, Niels Bohr Institute, University of Copenhagen, Juliane Maries Vej 30, DK-2100 Copenhagen \O, Denmark}
\altaffiltext{20}{\ignore{Steward}Steward Observatory, University of Arizona, 933 North Cherry Avenue, Tucson, AZ 85721, USA}
\altaffiltext{21}{\ignore{Caltech}Cahill Center for Astrophysics, California Institute of Technology, Pasadena, CA 91125, USA}
\altaffiltext{22}{\ignore{PSUExoP}Center for Exoplanets and Habitable Worlds, The Pennsylvania State University, \ignore{525 Davey Lab, }University Park, PA 16802, USA}
\altaffiltext{23}{\ignore{Missouri}Department of Physics and Astronomy, University of Missouri, Columbia, MO 65211, USA}
\altaffiltext{24}{\ignore{UCR}Department of Astronomy, University of California, Riverside, CA 92521, USA}
\altaffiltext{25}{\ignore{UCBoulder}Department of Astrophysical and Planetary Sciences, University of Colorado, Boulder, CO 80309, USA}
\altaffiltext{26}{\ignore{UCSC}Department of Astronomy and Astrophysics, University of California Santa Cruz, 1156 High Street, Santa Cruz, CA 95064, USA}
\altaffiltext{27}{\ignore{GeorgiaTech}Center for Relativistic Astrophysics, Georgia Institute of Technology, Atlanta, GA 30332, USA}
\altaffiltext{28}{\ignore{Washington}Department of Astronomy, University of Washington, Box 351580, Seattle, WA 98195, USA}
\altaffiltext{29}{\ignore{Lick}Lick Observatory, P.O.\ Box 85, Mt. Hamilton, CA 95140, USA}
\altaffiltext{30}{\ignore{JPL}Jet Propulsion Laboratory, California Institute of Technology, 4800 Oak Grove Drive, Pasadena, CA 91109, USA}
\footnotetext[31]{NASA Postdoctoral Program Fellow}
\altaffiltext{32}{\ignore{UCLA}Department of Physics and Astronomy, University of California, Los Angeles, CA 90095, USA}
\altaffiltext{33}{\ignore{SCIPP}Santa Cruz Institute for Particle Physics and Department of Physics, University of California, Santa Cruz, CA 95064, USA}
\altaffiltext{34}{\ignore{Stanford}Department of Physics, Stanford University, 382 Via Pueblo Mall, Stanford, CA 94305, USA}
\altaffiltext{35}{\ignore{Kavli}Kavli Institute for Particle Astrophysics and Cosmology, Stanford University, \ignore{452 Lomita Mall, }Stanford, CA 94305, USA}
\altaffiltext{36}{\ignore{SLAC}SLAC National Accelerator Laboratory, 2575 Sand Hill Road, Menlo Park, CA 94025, USA}
\altaffiltext{37}{\ignore{Melbourne}School of Physics, University of Melbourne, Parkville, VIC 3010, Australia}
\altaffiltext{38}{\ignore{Columbia}Department of Astronomy, Columbia University, 550 W120th Street, New York, NY 10027, USA}
\altaffiltext{39}{\ignore{MaxPlanck}Max Planck Institut f\"{u}r Astronomie, K\"{o}nigstuhl 17, D--69117 Heidelberg, Germany} 
\altaffiltext{40}{\ignore{NavalAcademy}Physics Department, United States Naval Academy, Annapolis, MD 21403, USA}
\altaffiltext{41}{\ignore{UCSB}Department of Physics, University of California, Santa Barbara, CA 93106, USA}
\altaffiltext{42}{\ignore{CfA}Harvard-Smithsonian Center for Astrophysics, 60 Garden Street, Cambridge, MA 02138, USA}
\footnotetext[43]{Einstein Fellow}
\footnotetext[44]{Sagan Fellow}
\altaffiltext{45}{\ignore{Michigan}Department of Astronomy, University of Michigan, 500 Church Street, Ann Arbor, MI 48109, USA}
\altaffiltext{46}{\ignore{Vale}Instituto de F\'{\i}sica, Universidade Federal do Rio do Sul, Campus do Vale, Porto Alegre, Brazil}
\altaffiltext{47}{\ignore{Carnegie}Carnegie Observatories, 813 Santa Barbara Street, Pasadena, CA 91101, USA}
\footnotetext[48]{Carnegie-Princeton Fellow, Hubble Fellow}
\altaffiltext{49}{\ignore{Vanderbilt}Department of Physics and Astronomy, Vanderbilt University, 6301 Stevenson Circle, Nashville, TN 37235, USA}
\altaffiltext{50}{\ignore{Millennium}Millennium Institute of Astrophysics, Santiago, Chile}
\altaffiltext{51}{\ignore{Mackenna}Instituto de Astrof\'isica, Pontificia Universidad Cat\'olica de Chile, Vicu\~na Mackenna 4860, Santiago, Chile}
\footnotetext[52]{Packard Fellow}
\altaffiltext{53}{\ignore{Valapaiso}Instituto de F\'{\i}sica y Astronom\'{\i}a, Facultad de Ciencias, Universidad de Valpara\'{\i}so, Gran Bretana N 1111, Playa Ancha, Valpara\'{\i}so, Chile}
\altaffiltext{54}{\ignore{Fermi}Osservatorio Astrofisico di Arcetri, largo E. Fermi 5, 50125, Firenze, Italy}
\altaffiltext{55}{\ignore{Crimean}Crimean Astrophysical Observatory, P/O Nauchny, Crimea 298409, Russia}
\altaffiltext{56}{\ignore{Fountainwood}Fountainwood Observatory, Department of Physics FJS 149, Southwestern University, 1011 E. University Ave., Georgetown, TX 78626, USA}
\altaffiltext{57}{\ignore{PSUphysics} Department of Physics, 104 Davey Laboratory, The Pennsylvania State University, University Park, PA 16802, USA}
\altaffiltext{58}{\ignore{Mullard}Mullard Space Science Laboratory, University College London, Holmbury St. Mary, Dorking, Surrey RH5 6NT, UK}
\altaffiltext{59}{\ignore{Wayne}Department of Physics and Astronomy, Wayne State University, 666 W. Hancock St, Detroit, MI 48201, USA}
\altaffiltext{60}{\ignore{WestKentucky}Department of Physics and Astronomy, Western Kentucky University, 1906 College Heights Blvd \#11077, Bowling Green, KY 42101, USA}
\altaffiltext{61}{\ignore{SUPA}SUPA Physics and Astronomy, University of St. Andrews, Fife, KY16 9SS Scotland, UK}
\altaffiltext{62}{\ignore{Ohio}Department of Physics and Astronomy, Ohio University, Athens, OH 45701, USA}
\altaffiltext{63}{\ignore{Worcester}Department of Earth, Environment and Physics, Worcester State University, \ignore{486 Chandler Street, }Worcester, MA 01602, USA}
\altaffiltext{64}{\ignore{Maryland}Department of Astronomy, University of Maryland, College Park, MD 20742, USA}
\altaffiltext{65}{\ignore{Pulkovo}Pulkovo Observatory, 196140 St.\ Petersburg, Russia}
\altaffiltext{66}{\ignore{UKent}Department of Physics and Astronomy, The University of Kentucky, Lexington, KY 40506, USA}
\altaffiltext{67}{\ignore{SDSU}Department of Astronomy, San Diego State University, San Diego, CA 92182, USA}
\altaffiltext{68}{\ignore{Goddard}Astrophysics Science Division, NASA Goddard Space Flight Center, Mail Code 661, Greenbelt, MD 20771, USA}
\altaffiltext{69}{\ignore{Canarias}Instituto de Astrof\'{\i}sica de Canarias, 38200 La Laguna, Tenerife, Spain}
\altaffiltext{70}{\ignore{Laguna}Departamento de Astrof\'{\i}sica, Universidad de La Laguna, E-38206 La Laguna, Tenerife, Spain}
\altaffiltext{71}{\ignore{GRANTECAN}Gran Telescopio Canarias (GRANTECAN), 38205 San Crist\'{o}bal de La Laguna, Tenerife, Spain}
\altaffiltext{72}{Spectral Sciences Inc., 4 Fourth Ave., Burlington, MA 01803, USA}
\altaffiltext{73}{Eureka Scientific Inc., 2452 Delmer St. Suite 100, Oakland, CA 94602, USA}
\altaffiltext{74}{\ignore{Morehead}Space Science Center, Morehead State University, 235 Martindale Dr., Morehead, KY 40351, USA}
\altaffiltext{75}{\ignore{York}Department of Physics and Astronomy, York University, Toronto, ON M3J 1P3, Canada}
\altaffiltext{76}{\ignore{Seoul}Astronomy Program, Department of Physics \& Astronomy, Seoul National University, Seoul, Republic of Korea}
\altaffiltext{77}{\ignore{BYU}Department of Physics and Astronomy, N283 ESC, Brigham Young University, Provo, UT 84602, USA}
\altaffiltext{78}{\ignore{SRON}SRON Netherlands Institute for Space Research, Sorbonnelaan 2, 3584 CA Utrecht, The Netherlands}
\altaffiltext{79}{\ignore{Utrecht}Department of Physics and Astronomy, Univeristeit Utrecht, P.O. Box 80000, 3508 Utrecht, The Netherlands}
\altaffiltext{80}{\ignore{Leiden}Leiden Observatory, Leiden University, PO Box 9513, 2300 RA Leiden, The Netherlands}
\altaffiltext{81}{\ignore{TelAviv}School of Physics and Astronomy, Raymond and Beverly Sackler Faculty of Exact Sciences, Tel Aviv University, Tel Aviv 69978, Israel}
\altaffiltext{82}{\ignore{Technion}Physics Department, Technion, Haifa 32000, Israel}
\altaffiltext{83}{\ignore{Korea}Korea Astronomy and Space Science Institute, Republic of Korea}
\altaffiltext{84}{\ignore{}Departamento de Astronomia, Universidad de Chile, Camino del Observatorio 1515, Santiago, Chile}
\altaffiltext{85}{\ignore{Southampton}University of Southampton, Highfield, Southampton, SO17 1BJ, UK}
\altaffiltext{86}{\ignore{DiSAT}DiSAT, Universita dell'Insubria, via Valleggio 11, 22100, Como, Italy}
\altaffiltext{87}{\ignore{Crete}Department of Physics and Institute of Theoretical and Computational Physics, University of Crete, GR-71003 Heraklion, Greece}
\altaffiltext{88}{\ignore{IESL}IESL, Foundation for Research and Technology, GR-71110 Heraklion, Greece}
\altaffiltext{89}{\ignore{Haifa}Department of Physics, Faculty of Natural Sciences, University of Haifa, Haifa 31905, Israel}
\altaffiltext{90}{\ignore{LCOGT}Las Cumbres Observatory Global Telescope Network, 6740 Cortona Drive, Suite 102, Goleta, CA 93117, USA}
\altaffiltext{91}{\ignore{Amsterdam}Astronomical Institute `Anton Pannekoek,' University of Amsterdam, Postbus 94249, NL-1090 GE Amsterdam, The Netherlands}
\altaffiltext{92}{\ignore{Bath}University of Bath, Department of Physics, Claverton Down, BA2 7AY, Bath, UK}
\altaffiltext{93}{\ignore{CMU}Department of Physics, Carnegie Mellon University, 5000 Forbes Avenue, Pittsburgh, PA 15213, USA}

\begin{abstract}
We present the results of an optical spectroscopic monitoring program targeting \nagn as part of a larger multi-wavelength reverberation mapping campaign. The campaign spanned six months and achieved an almost daily cadence with observations from five ground-based telescopes. The \hbeta and \heiiopt broad emission-line light curves lag that of the 5100~\angstrom optical continuum by $4.17^{+0.36}_{-0.36}$ days and $0.79^{+0.35}_{-0.34}$ days, respectively. The \hbeta lag relative to the 1158~\angstrom ultraviolet continuum light curve measured by the \textit{Hubble Space Telescope} is roughly $\sim$50\% longer than that measured against the optical continuum, and the lag difference is consistent with the observed lag between the optical and ultraviolet continua. This suggests that the characteristic radius of the broad-line region is $\sim$50\% larger than the value inferred from optical data alone. We also measured velocity-resolved emission-line lags for \hbeta and found a complex velocity-lag structure with shorter lags in the line wings, indicative of a broad-line region dominated by Keplerian motion. The responses of both the \hbeta and \heii emission lines to the driving continuum changed significantly halfway through the campaign, a phenomenon also observed for \civ, \lya, \ion{He}{2}(+\ion{O}{3}]), and \ion{Si}{4}(+\ion{O}{4}]) during the same monitoring period. Finally, given the optical luminosity of \nagn during our campaign, the measured \hbeta lag is a factor of five shorter than the expected value implied by the \rl relation based on the past behavior of NGC 5548.
\end{abstract}

%====================================================================================================

\section{Introduction}

Broad emission lines are among the most striking features of quasars and active galactic nuclei (AGNs). These Doppler-broadened lines are emitted by gas occupying the broad-line region (BLR), which is located within several light-days to light-months of the central supermassive black hole \citep[SMBH; e.g.,][]{antonuccicohen1983,clavel1991,peterson1998,peterson2004,bentz2009_lamp,grier2013}. The geometry and kinematics of the BLR play a significant role in AGN research because these properties can be used to infer the mass of the central black hole \citep[e.g.,][]{gaskellsparke1986,clavel1991,kaspi2000,denney2006,denney2010,pancoast2014}. Additionally, it is possible that infalling BLR gas may fuel SMBH accretion \citep[e.g.,][]{peterson2006,gaskellgoosmann2016} and outflowing gas may be part of disk winds that carry away angular momentum from the disk and provide energy and momentum feedback to the host galaxy \citep[e.g.,][]{emmering1992,murraychiang1997,kollatschny2003,leighlymoore2004}. Understanding the dynamical state and physical conditions of gas in the BLR is of key importance in completing our understanding of the AGN phenomenon.

Owing to its small angular size, the BLR is currently impossible to resolve spatially even for the closest AGNs. An alternative method to study this region is to resolve it in the time domain using reverberation mapping (RM), a technique that leverages the variable nature of quasars and Seyferts \citep{blandfordmckee,peterson1993,peterson2014}. AGNs exhibit stochastic flux variations, possibly because of inhomogeneous accretion and thermal fluctuations in the accretion disk \citep{czerny1999,collierpeterson2001,czerny2003,kelly2009,kozlowski2010,macleod2010}. Photons from the central engine ionize the BLR gas, which then echoes continuum-flux variations with a light-travel time lag, $\tau$. The emission-line flux $L(v_r,t)$ at time $t$ and line-of-sight velocity $v_\mathrm{r}$ is related to the ionizing continuum by 

\begin{equation}
L(v_r,t) = \int_{0}^{\infty} \Psi(v_r,\tau) C(t-\tau) d\tau,
\label{eq:transferfunction}
\end{equation}

\noindent where $C(t-\tau)$ is the continuum emission at an earlier time $t-\tau$, and $\Psi(v,\tau)$ is the transfer function that maps the continuum light curve to the time-variable line profile \citep{blandfordmckee}.

The transfer function---also known as the velocity-delay map---encodes important information about the BLR's geometry and kinematics. There has been tremendous effort by many groups to recover velocity-delay maps \citep{rosenblatt1990,horne1991,krolik1991,ulrichhorne1996,bentz2010_arp151,caramel2011,liyanrong2013,grier2013,pancoast2014,skielboe2015} and velocity-resolved line lags \citep[e.g.,][]{kollatschny2003,bentz2009_lamp,denney2010,barth_zw229,pudu_velres}. In order to obtain $\Psi(v,\tau)$, RM campaigns must have a combination of high cadence, long duration, high photometric precision, and high signal-to-noise ratios (SNR), which is often not achievable by ground-based programs. More typically, RM campaigns are able to only measure the mean emission-line lag $\tau$, which represents the response-weighted mean light-travel time from the ionizing continuum to the BLR.

Assuming that the broad-line width is a result of the virialized motion of gas within the black hole's potential well, the emission-line lag and gas velocity dispersion inferred from the line width ($\Delta V$) can be used to infer the black hole (BH) mass using

\begin{equation}
M_{\mathrm{BH}} = f \frac{c\tau\Delta V^2}{G}.
\label{eq:rm}
\end{equation}

\noindent Here, $c\tau=\rblr$ is the characteristic radius of the BLR, and \textit{f} is a dimensionless calibration factor of order unity that accounts for the unknown BLR geometry and kinematics. Ground-based RM campaigns have produced BH mass measurements for $\sim$60 local AGNs to date (see \citealt{agndatabase} for references and a recent compilation). RM is also starting to be used for objects at cosmological distances \citep{kaspi2007,king2015,shen2016} with the aims of studying the UV continuum and emission lines and calibrating BH masses at high redshifts.

The ionizing continuum is emitted at wavelengths $\textless$ 912~\angstrom and is generally unobservable due to the Lyman limit of the host galaxy. Given this limitation, the far-ultraviolet (UV) continuum at $\lambda \approx 1100$--1500~\angstrom should be used to derive emission-line lags because it is close in wavelength to the ionizing continuum and should therefore serve as an accurate proxy. However, wavelengths shorter than $\sim$3200~\angstrom are inaccessible from the ground, so the rest-frame optical continuum is often used as a proxy for the ionizing source in low-redshift AGNs. Although the far-UV and optical continua have been shown to vary almost simultaneously in some cases \citep[e.g.,][]{clavel1991,reichert1994,korista1995,wanders1997}, more recent high-cadence studies have found that the optical continuum can lag the UV continuum by up to a few days \citep{collier1998,sergeev2005,mchardy2014,shappee2014,stormII,stormIII}. This can significantly affect the measured broad-line lag if the BLR has a characteristic radius on the order of light days. The variable optical continuum has also been shown to have smoother features and smaller amplitudes than its UV counterpart \citep[e.g.,][]{peterson1991,dietrich1993,stirpe1994,santos-lleo1997,dietrich1998,shappee2014,stormIII}. These differences between the UV and optical continua suggest that the optical continuum is not fully interchangeable with the ionizing source for determining reverberation lags.

Furthermore, a long-standing assumption in RM is that the source of the ionizing photons in a typical Seyfert galaxy is physically much smaller than the BLR \citep[about a factor of 100; e.g.,][]{peterson1993,petersonhorne2004}. This assumption implies that the disk size can be neglected when determining \rblr from RM data. However, \citet{stormIII} have shown that the optically emitting portion of the accretion disk has a lag similar to that of the inner portion of the BLR. If we assume a model in which the measured lags are purely dependent on the radial distance from the ionizing source, then the emission-line lags measured using the optical continuum may significantly underestimate the BLR characteristic radius. Since most RM campaigns use only optical data, it is imperative that we understand the systematic effects of using the optical rather than the UV continuum in RM studies and the relevant implications for BH mass estimates.

To this end, we present the results of a six-month ground-based RM program monitoring the galaxy \nagn (redshift $z=0.0172$). This paper is the fifth in a series describing results from the AGN Space Telescope and Optical Reverberation Mapping (AGN STORM) campaign, the most intensive multi-wavelength AGN monitoring program to date. The campaign is centered around 171 epochs of daily cadence observations using the Cosmic Origins Spectrograph on the \textit{Hubble Space Telescope} (\hst). Concurrent with the \hst program were four months of \swift observations and six months of ground-based photometric and spectroscopic observations. First results of the \hst, \swift, and ground-based photometry programs were presented by \citet{stormI}, \citet{stormII}, and \citet{stormIII} (Papers I--III, respectively). \citet{stormIV} (Paper IV) explore the anomalous behavior of the UV continuum and broad emission-line light curves observed during a portion of this campaign. This paper focuses on the ground-based spectroscopic data and emission-line analysis.
 
\nagn is one of the best-studied Seyfert galaxies and has been the subject of many past RM programs. Most notably, it was the target of a 13-year campaign carried out by the AGN Watch consortium \citep[][and references therein]{agnwatchXVI}, which was initially designed to support UV monitoring of \nagn carried out by the \textit{International Ultraviolet Explorer} \citep[\textit{IUE;}][]{clavel1991}. Individual years of this campaign achieved median sampling cadences of 1--3 days for spectroscopic observations. Subsequently, \nagn was monitored in programs described by \citet{bentz2007}, \citet{denney2009}, \citet{bentz2009_lamp}, and De Rosa et al. (AGN12, results in preparation) with campaign durations of 40 days, 135 days, 64 days, and 120 days (respectively), and each with a median sampling cadence of $\sim1$ day. A more recent RM program described by \citet{kxlu2016} monitored this AGN for 180 days with a median spectroscopic sampling of $~\sim3$ days. The 2014 AGN STORM campaign's combination of daily cadence, six-month duration, and multi-wavelength coverage makes it the most intensive RM campaign ever conducted.

There are two primary goals of the present work. The first is to compare the \hbeta emission-line lag measured against simultaneously observed far-UV and optical continua in order to understand the effects of substituting the optical continuum for the ionizing continuum in reverberation measurements. The second goal is to examine in detail the responses of the optical emission lines to continuum variations and compare them to those of the UV lines, which will provide a more complete picture of the structure and kinematics of the BLR than previous studies that used only optical data.

We describe the spectroscopic observations and reductions in Section 2. Section 3 details our procedures for flux and light-curve measurements. In Section 4, we present our analysis of emission-line lags, line responses, line profiles, and BH mass measurements. We discuss the implications of our results and compare our measurements to those from previous campaigns in Section 5. Section 6 summarizes our findings. We quote wavelengths in the rest frame of \nagn unless otherwise stated.

%====================================================================================================

\section{Observations and Data Reduction}

Spectroscopic data were obtained from five telescopes: the McGraw-Hill 1.3-m telescope at the MDM Observatory, the Shane 3-m telescope at the Lick Observatory, the 1.22-m Galileo telescope at the Asiago Astrophysical Observatory, the 3.5-m telescope at Apache Point Observatory (APO), and the 2.3-m telescope at the Wyoming Infrared Observatory (WIRO). Observations at MDM were carried out with a slit width of 5\arcsec~oriented in the north-south direction, and spectra at the other telescopes were taken with a 5\arcsec-wide slit oriented at the parallactic angle \citep{filippenko82}. The optical spectroscopic monitoring began on 2014 January 4 (UT dates are used throughout this paper) and continued through 2014 July 6 with approximately daily cadence. 

\begin{figure}
\centering
\includegraphics[angle=0,scale=0.53,trim={0.0cm 0.2cm 0.0cm 0.6cm},clip=true]{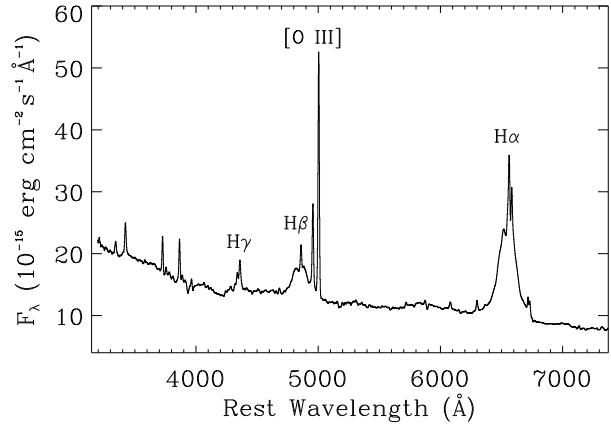}
\caption{Mean spectrum of \nagn from the Asiago dataset, which includes 21 epochs of spectra with spectral resolution of 1.0 \angstrom\perpix and has a median SNR of 160. Labeled are the \heii \lam4686 \hbeta \lam4861, and \oiii \lam\lam 4959, 5007 emission lines.}
\label{fig:asiagomeanspec}
\end{figure}

Table~\ref{tab:telinfo} lists the properties of the telescopes and instruments used to obtain spectroscopic data, and Figure~\ref{fig:asiagomeanspec} shows the mean spectrum constructed using data from Asiago, which obtained the only spectra that cover the full optical wavelength range. MDM contributed the largest number of spectra with 143 epochs. The 35 epochs of Lick spectra were obtained by several groups of observers who used slightly different setups and calibrations. The Kast spectrograph \citep{millerstone93} at Lick Observatory has red-side and blue-side cameras, but since the red-side setup was very different for each group, we present only the blue-side data here. Asiago, APO, and WIRO contributed 21, 13, and 6 epochs of spectra, respectively. Our analysis focuses primarily on the MDM dataset for homogeneity.

%\begin{landscape}
\begin{deluxetable*}{llccccccc}
  \tablecaption{Instrument characteristics and data-reduction parameters for all telescopes}
  \tablewidth{0pt}
  \tablecolumns{8}
  \tablehead{
    \colhead{Telescope} &
    \colhead{Instrument} &
    \colhead{Number of} &
    \colhead{Median} &
    \colhead{Wavelength} &
    \colhead{Wavelength} &
    \colhead{Pixel} &
    \colhead{Median} &
    \colhead{\oiii} \\
    \colhead{} &
    \colhead{} &
    \colhead{Epochs} &
    \colhead{Seeing} &
    \colhead{Dispersion} &
    \colhead{Coverage} &
    \colhead{Scale} &
    \colhead{SNR} &
    \colhead{$F_\mathrm{var}$} \\
    \colhead{} &
    \colhead{} &
    \colhead{} &
    \colhead{(\arcsec)} &
    \colhead{(\angstrom\perpix)} &
    \colhead{(\angstrom)} &
    \colhead{(\arcsec\xspace\perpix)} &
    \colhead{} &
    \colhead{(\%)}
}
\startdata
    MDM     & Boller \& Chivens CCD Spectrograph & 143  & 1.7 & 1.25  &  4225$-$5775  &  0.75  & 118 &  0.62 \\
    Lick    & Kast Double Spectrograph           & 35   & 1.5 & 1.02  &  3460$-$5500  &  0.43  & 194 &  0.32 \\
    Asiago  & Boller \& Chivens CCD Spectrograph & 21   & 4.0 & 1.00  &  3250$-$7920  &  1.00  & 160 &  0.27 \\
    APO     & Dual Imaging Spectrograph          & 13   & 1.4 & 1.00  &  4180$-$5400  &  0.41  & 160 &  0.28 \\
    WIRO    & WIRO Long Slit Spectrograph        & 6    & 2.1 & 0.74  &  5599$-$4399  &  0.52  & 217 &  0.47

\enddata
\tablecomments{The wavelength coverage for Lick refers to only the Kast blue-side camera. The SNR value refers to the median SNR per pixel over the rest wavelength range 5070$-$5130 \angstrom. The \oiii $F_\mathrm{var}$ is the amount of residual variations in the \oiii light curve after spectral scaling and gives an indication of the flux scaling accuracy.}
\label{tab:telinfo}
\end{deluxetable*}
%\end{landscape}

%--------FIGURE--------%
%----------------------%

Data-reduction procedures included bias subtraction, flat fielding, and cosmic ray removal using the L.A. Cosmic routine \citep{vandokkum}. The one-dimensional spectra were extracted from a 15\arcsec-wide region centered on the AGN and with consistent background sky apertures for all observations. We used optimally-weighted extractions for the stellar spectra \citep{horne1986} but unweighted extractions for the AGN spectra. This is because the optimal extraction method requires the spatial profile of the target to be a smooth function of wavelength, and tends to truncate the peaks of strong emission lines such as \oiii that have different spatial extents from the surrounding continuum.

The data were wavelength calibrated using night sky lines and flux calibrated using standard stars. Our most frequently used flux standard stars were Feige 34, BD 332642, and HZ 44. For nights when multiple exposures were taken, we aligned the flux-calibrated one-dimensional spectra by applying small wavelength shifts to each spectrum before combining them. We do not expect significant differential atmospheric refraction \citep{filippenko82} because of the large slit width use for our observations. 

For the MDM data, the first 133 epochs were flux calibrated using Feige 34, while the last 10 epochs, taken from 2014 June 20 to 2014 June 30, were flux calibrated with BD 332642. This caused spurious changes in the shape of some emission-line features, so we use only the first 133 MDM epochs for our present analysis. 

\subsection{Spectral Flux Calibrations}

To place the instrumental fluxes on an absolute flux scale, we measured the narrow \oiii \lam5007 line flux from spectra taken under photometric conditions and scaled all other nightly spectra to have the same \oiii flux. There were 21 epochs identified as having been observed under photometric conditions by the MDM observers. We determined the flux of the \oiii line ($\lambda_\mathrm{observed}=5093$~\angstrom) by first subtracting a linear fit to continuum windows on either side of the line, then integrating over a fixed wavelength range.  We used the rest-frame wavelength ranges 4976.5$-$4948.0 \angstrom and 5027.7$-$5031.6 \angstrom to fit the continuum and integrated over the range 4980.5$-$5026.7 \angstrom for the line flux. The 2$\sigma$ outliers from this set of \oiii flux measurements were discarded, the mean was recomputed, and this process was repeated until there were no more 2$\sigma$ outliers, which resulted in a total of 16 final photometric spectra. The mean spectrum of these 16 epochs has an \oiii \lam5007 line flux of $(5.01 \pm 0.11) \times 10^{-13}$ \ergscm, which represents our best estimate of the true \oiii flux for \nagn during this campaign and is not expected to vary over a six-month period. For comparison, \citet{peterson13_nlvar} found the \oiii flux in \nagn to be $(4.77 \pm 0.14) \times 10^{-13}$ \ergscm in their 2012 monitoring campaign, and the difference is within the range of total \oiii variability observed for \nagn over the course of 21 years \citep[see][]{peterson13_nlvar}.

In addition to the intrinsic variability of the AGN, many other factors contribute to nightly variations in the spectra. These include changes in transparency due to clouds, changes in seeing conditions, inconsistent instrument focus, and miscentering of the AGN in the slit during observations. We used the flux scaling method described by \citet{vgw} to align the nightly spectra and place them on a consistent flux scale. For each spectrum in the dataset, the algorithm looks for a combination of wavelength shift, multiplicative scale factor, and Gaussian kernel convolution that minimizes the residual between each individual spectrum and a reference spectrum over a region containing the narrow \oiii line. 

We constructed a separate reference spectrum for each telescope by averaging the highest SNR spectra in each dataset, then broadened the reference spectrum so that the \oiii line width matches the broadest \oiii line width in the dataset. This extra broadening of the reference spectrum helps to reduce the \oiii residuals from spectral scaling \citep{fausnaugh2016_scaling}. We then scaled each spectrum to have the same \oiii flux as the photometrically calibrated mean MDM spectrum. This brings all spectra to a common flux scale after spectral scaling.

To assess the accuracy of spectral scaling, we estimated the intrinsic fractional variability of the residual \oiii \lam5007 light curve after correcting for random measurement errors,

\begin{equation}
F_{\mathrm{var}} = \frac{\sqrt{\sigma^2 - \langle\delta^2\rangle}}{\langle f\rangle},
\label{eq:fvar}
\end{equation}

\noindent where $\sigma^2$ is the \oiii flux variance, $\langle \delta^2 \rangle$ is the mean-square value of the measurement uncertainties determined from the nightly error spectra produced by the data reduction pipeline, and $\langle f \rangle$ is the unweighted mean flux. The $F_\mathrm{var}$ for the \oiii \lam5007 light curve gives a good estimate of the residual flux-scaling errors \citep{barthbentz2016}, and the value for each telescope is listed in the last column of Table~\ref{tab:telinfo}. We found $F_\mathrm{var}$ to be between 0.27\% and 0.62\% for all telescopes, which means there is an additional scatter of less than 1\% in the \oiii light curve above the measurement errors. These $F_\mathrm{var}$ values are consistent with or better than the best values typically obtained in ground-based campaigns. For example, \citet{barth2015} found $F_\mathrm{var}$ values ranging from 0.5\% to 3.3\% for individual AGNs in the 2011 Lick AGN Monitoring Project.

\begin{figure}
\centering
\includegraphics[angle=0,scale=0.58,trim={0.0cm 0.1cm 0.0cm 0.4cm},clip=true]{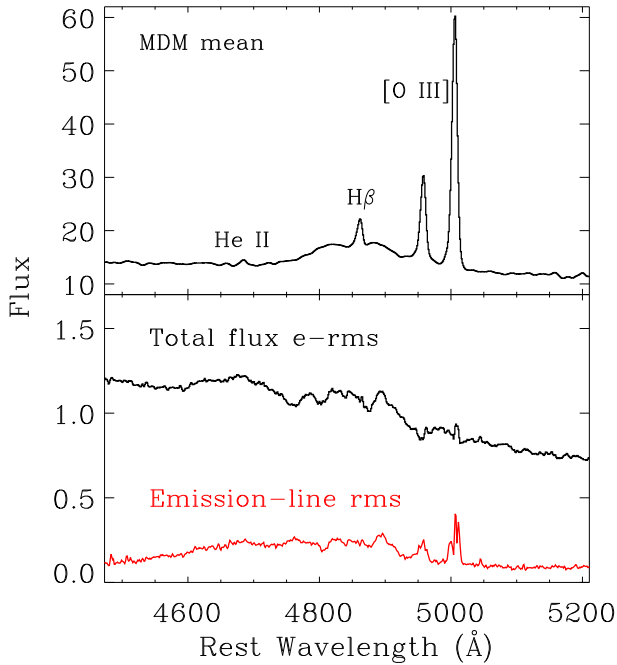}
\caption{The mean and excess rms (Eq.~\ref{eq:excessvariance}) spectra from the MDM dataset are shown in black, and the rms spectrum with the AGN and stellar continuum removed is shown in red (see Section 3.1).}
\label{fig:mean_rms_full}
\end{figure}

Figure~\ref{fig:mean_rms_full} shows (in black) the mean and root-mean-square (rms) residual spectra for the MDM dataset. The rms spectrum indicates the degree of variability at each wavelength over the course of the campaign. Both the broad \hbeta and \heiiopt emission lines exhibit strong variations, and the \hbeta rms profile appears to have multiple peaks. Traditionally, the rms spectrum is constructed such that the value at each wavelength is taken to be the standard deviation of fluxes from all epochs, but this does not take into account Poisson or detector noise, which may bias the rms profile by a small amount \citep{barth2015}. \citet{park2012_lamp} suggest using the SNR for each spectrum as the weight for that spectrum in calculating the rms, or using a maximum-likelihood method to obtain the rms. We adopt a simpler approach that uses the excess variance as a way to exclude variations that are not intrinsic to the AGN. This ``excess rms'' value at each wavelength is defined as

%--------EQUATION-------%
\begin{equation}
\mathrm{e\text{-}rms}_{\lambda} = \sqrt{\frac{1}{N-1}\displaystyle\sum\limits_{i=1}^{N} [(F_{\lambda,i} - \langle F_{\lambda}\rangle)^{2} - \delta^{2}_{\lambda,i}]},
\label{eq:excessvariance}
\end{equation}
%-----------------------%

\noindent where $N$ is the total number of spectra in the dataset, $\langle F_{\lambda}\rangle$ is the mean flux at each wavelength, and $F_{\lambda,i}$ and $\delta_{\lambda,i}$ are the wavelength-specific fluxes and associated measurement uncertainties from individual epochs, respectively. This method estimates the degree of variability above what is expected given the measurement uncertainties and pixel-to-pixel noise. 

%--------FIGURE--------%
%----------------------%

%~~~~~~~~~~~~~~~~~~~~~~~~~~~~~~~~~~~~~~~~~~~~~~~~~~~~~~~~~~~~~~~~~~~~~~~~~~~~~~~~~~~~~~~~~~~~~~~~
\section{Spectroscopic Flux Measurements}

The 5100~\angstrom continuum flux density was determined by averaging the flux over the rest-frame wavelength range 5070$-$5130\angstrom. The \hbeta line fluxes were measured from the scaled spectra using the same method as for \oiii \lam5007, where we subtracted a linear fit to the surrounding continuum (wavelength windows 4483.0$-$4542.0 \angstrom and 5033.5$-$5092.5 \angstrom) and integrated across the line profile (4748.4$-$4945.1 \angstrom). The uncertainty in each measurement is a combination of Poisson noise and residuals from spectral scaling. We computed the spectral scaling uncertainty by multiplying each flux measurement by the \oiii $F_\mathrm{var}$ value for that dataset, then adding this value in quadrature to the Poisson noise to obtain the final flux uncertainty for each measurement. There is an additional source of spectral scaling uncertainty from slight differences in the overall spectral shape from night to night. This effect is likely small for \hbeta because it is very close to the \oiii \lam5007 line that anchors the spectral scaling.

Spectrophotometric calibrations of the reference spectra, as described in the previous section, converted all instrumental fluxes to absolute fluxes, which means that measurements from all telescopes should now be on the same flux scale. However, light curves from different observing sites may be offset from each other owing to aperture effects \citep{peterson1995,peterson1999}. While our observations were standardized to have the same 5\arcsec~$\times$ 15\arcsec~aperture size, significant differences in image quality between observing sites could still cause flux offsets.

To intercalibrate the \hbeta light curves, we used data points from each non-MDM telescope ($F_\mathrm{\hbeta,t}$) that are nearly contemporaneous with MDM observations ($F_\mathrm{\hbeta,MDM}$) and performed a least-squares fit to the equation

\begin{equation}
F_\mathrm{\hbeta,MDM} = \phi F_\mathrm{\hbeta,t}
\end{equation}

\noindent to find the scale factor $\phi$ that puts each line light curve on the same flux scale as the MDM data. For the continuum intercalibration, we also include an additive shift $G$ to account for the differences in the host-galaxy flux admitted by different apertures: 

\begin{equation}
F_\mathrm{5100,MDM} = \phi F_\mathrm{5100,t} + G.
\end{equation}

\begin{figure*}
\centering
\includegraphics[angle=0,scale=0.80]{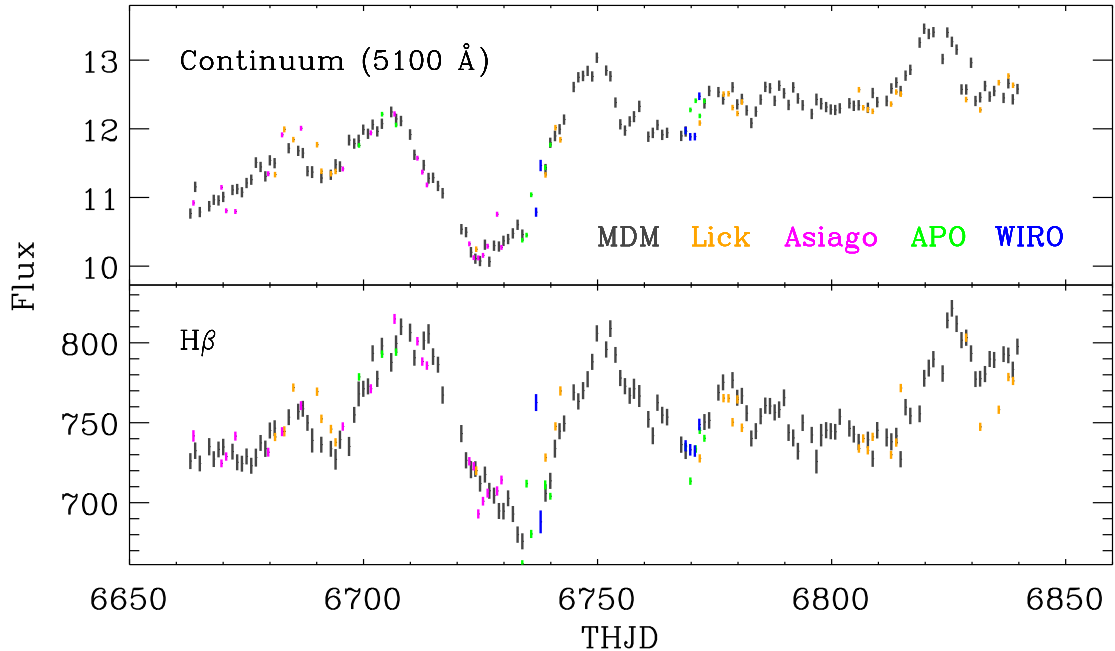}
\caption{Continuum (\fluxunitlambda) and \hbeta (\fluxunit) light curves (THJD = HJD $-$ 2,450,000). The Lick, APO, Asiago, and WIRO light curves were scaled and shifted to match the MDM light curve, which has the longest temporal coverage and highest sampling cadence. The plotted uncertainties include Poisson noise and the normalized excess variance of the \oiii light curve (Section 2.1.)}
\label{fig:cont_hbeta_alltel}
\end{figure*}

\noindent The scale factors for the Lick, Asiago, APO, and WIRO light curves are $\phi = [0.961,~0.963,~1.037,~0.918]$, and the shift constants are $G = $[$-0.155$, $-0.640$, $-0.041$, $0.024$] in units of \fluxunitlambda. The combined continuum and \hbeta light curves are shown in Figure~\ref{fig:cont_hbeta_alltel} (THJD = HJD $-$ 2,450,000), and the 5100~\angstrom continuum and \hbeta fluxes are listed in Table~\ref{tab:fluxtable}.

%--------FIGURE--------%
%----------------------%

%-----TABLE-----%
%\begin{landscape}
\begin{deluxetable*}{clcccc}
%\tabletypesize{}
%\rotate
\tablewidth{0pt} %0pt is natural width
\tablecolumns{6}
\tablecaption{Flux measurements for continuum and emission lines.\label{tab:fluxtable}}
\tablehead{
 \colhead{HJD$-$2,450,000}  & 
 \colhead{Telescope}  & 
 \colhead{$F_{5100}$}  & 
 \colhead{$F_\mathrm{\hbeta}$}  & 
 \colhead{$F_\mathrm{\hbeta,SD}$}  & 
 \colhead{$F_\mathrm{He~II,SD}$} 
}
\startdata

   6663.00 & MDM   & 10.766 $\pm$ 0.075 & 726.012 $\pm$ 4.985 & 710.187 $\pm$  4.897 &  21.720  $\pm$ 2.606 \\
   6663.65 & Asiago& 10.921 $\pm$ 0.040 & 741.771 $\pm$ 3.586 &         ---         &            ---        \\
   6664.03 & MDM   & 11.154 $\pm$ 0.075 & 732.511 $\pm$ 5.156 & 715.057 $\pm$  5.061 &  28.154  $\pm$ 3.222 \\
   6665.02 & MDM   & 10.788 $\pm$ 0.075 & 724.537 $\pm$ 4.946 & 709.473 $\pm$  4.860 &  25.135  $\pm$ 2.451 \\
   6667.02 & MDM   & 10.872 $\pm$ 0.076 & 735.001 $\pm$ 5.393 & 711.347 $\pm$  5.288 &  37.008  $\pm$ 3.964 

\enddata

\tablecomments{The 5100~\angstrom continuum flux density (\fluxunitlambda) includes contributions from both the AGN and the host galaxy. The \hbeta and \hbeta$_\mathrm{SD}$ fluxes were obtained using a linear continuum model and the spectral decomposition method, respectively. The \heii flux is based on the spectral decomposition model. All emission-line fluxes are in units of \fluxunit and include contributions from both broad- and narrow-line components. The full table is available in the online version.}
\end{deluxetable*}
%\end{landscape}

We attempted to measure the \heiiopt flux from the nightly spectra. However, this line is very weak and also heavily blended with the broad \hbeta, as shown in Figure~\ref{fig:mean_rms_full}. Thus, we were unable to obtain a \heii light curve using the linear interpolation method to remove the continuum.

\subsection{Spectral Decomposition}

To more accurately remove the continuum underlying the emission lines and to deblend the broad emission features from each other, we employed the spectral decomposition algorithm described by \citet{barth2015}. The components fitted in this procedure include narrow \oiii, broad and narrow \hbeta, broad and narrow \heii, \feii emission blends, the stellar continuum, and the AGN continuum. The host-galaxy starlight was modeled with an 11 Gyr, solar metallicity, single-burst spectrum from \citet{bruzualcharlot2003}. For the \feii model component, we tested three different templates from \citet{bg}, \citet{veron}, and \citet{kovacevic}. The \feii templates were broadened by convolution with a Gaussian kernel in velocity. The free fit parameters for \feii include the velocity shift relative to broad \hbeta, the broadening kernel width, and the flux normalization of the broadened template spectrum. The \citet{bg} and \citet{veron} templates are monolithic and require only one flux normalization parameter, whereas the \citet{kovacevic} template has five components that can vary independently in flux. The \citet{kovacevic} template achieves the best fit to the nightly spectra, presumably a result of the larger number of free fit parameters due to the multi-component \citet{kovacevic} template.

%--------FIGURE--------%
\begin{figure}
\centering
\includegraphics[angle=0,scale=0.56]{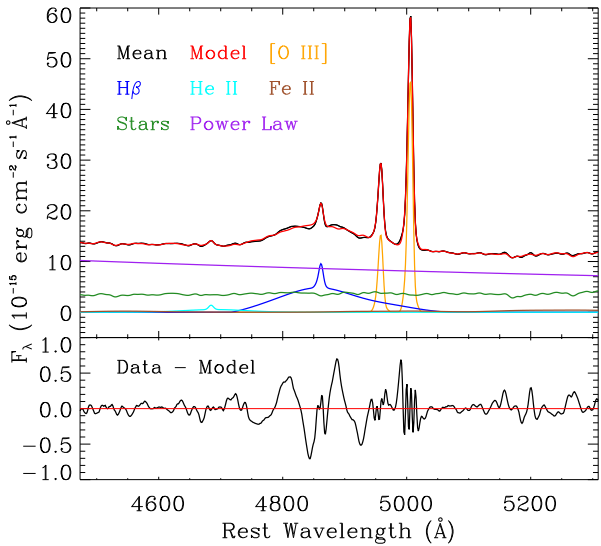}
\caption{\textit{Top:} Spectral decomposition components of the mean MDM spectrum. The red spectrum is the sum of all model components and traces the data (black spectrum) well over most of the spectral range. \textit{Bottom:} Residuals from the full model fit.}
\label{fig:specdecomp}
\end{figure}
%----------------------%

We made several modifications to the spectral fitting procedures used by \citet{barth2015}. First, because of the complex line profiles, we used sixth-order Gauss-Hermite functions \citep{vandermarel1993} to fit the broad and narrow \hbeta and narrow \oiii lines instead of fourth-order functions. Second, there is significant degeneracy between the weak \feii blend and the continuum flux in the nightly fits. Since the \feii fit is poorly constrained and sometimes varied drastically from night to night, the continuum model flux also varied significantly as a result, which in turn introduced noise to the broad \hbeta fit component. To address this issue, we constrained the \feii flux to lie within 10\% of the value from the fit to the mean spectrum \citep{barth_feii}. We also fixed the \feii redshift to that of the mean spectrum and constrained the \feii broadening kernel to be within 5\% of its value from the mean spectrum fit. The \hei \lam4922 and \lam5016 lines are very weak and are heavily blended with broad \hbeta, making it impossible to constrain their fit parameters. We therefore do not fit for these components in our model.

The broad \heiiopt component has very low amplitude compared to the other fit components and it is blended with the blue wing of broad \hbeta. It is also highly variable, as demonstrated by the broad bump in the rms spectrum. This made it difficult to fit the \heii broad-line profile accurately, and the width varied significantly from night to night when fitted as a free parameter. Since the \heii $\lambda1640$ and $\lambda4868$ lines are expected to form under the same physical conditions and should thus have similar widths, we used fits to the $\lambda1640$ line in concurrent \hst spectra to constrain the $\lambda4686$ line width.

The \heii $\lambda1640$ line was modeled with five Gaussian components (De Rosa et al., in prep.), and we took the three broadest components to represent the broad \heii $\lambda1640$ line profile. For each MDM spectrum, the \heii $\lambda4686$ broad-line full width at half-maximum intensity (FWHM) was allowed to vary within 3~\angstrom of the \heii $\lambda1640$ FWHM measured from the closest \hst epoch. The first 23 epochs from the MDM campaign do not have corresponding \hst spectra, so for each of these ``pre-\hst'' epochs, we found the three epochs from later in the campaign with the closest matching 5100~\angstrom continuum flux density. We then used the weighted mean of the broad \heii $\lambda1640$ widths from these three nights as the width constraint for the pre-\hst epoch, where the weight were determined by how closely the 5100~\angstrom fluxes of the later epochs matched that of the pre-\hst epoch. The \heii $\lambda1640$ line width was highly variable during the \hst campaign, and the model FWHM widths used to constrain the spectral decomposition have a mean of 48~\angstrom, with a minimum of 28~\angstrom and maximum of 59~\angstrom. 

We applied spectral decomposition to the data from all telescopes, but since the MDM dataset is the largest and has the highest data quality and consistency, we use this dataset for all subsequent analysis. Figure~\ref{fig:specdecomp} shows the fit components for the mean MDM spectrum, where the black spectrum is the data and the red spectrum is the sum of all the model components. The model does not fit the detailed structure of the broad \hbeta line well, especially in the line core. To prevent this from impacting our measured \hbeta fluxes, we subtracted all the other well-modeled fit components except the broad and narrow \hbeta components from the full spectrum, then obtained the \hbeta line flux by integrating over the same wavelength range used to measure the flux without spectral decomposition. The \heiiopt flux was taken to be the total flux in the broad- and narrow-line models for each night. The narrow \hbeta and \heiiopt line fluxes from fits to the mean spectrum are $48.4 \times 10^{-15}~\mathrm{erg~s}^{-1}~\mathrm{cm}^{-2}$ and $8.5 \times 10^{-15}~\mathrm{erg~s}^{-1}~\mathrm{cm}^{-2}$  (respectively), with uncertainties of $\sim 2$\% from the overall photometric scale of the data. The ratio of the narrow \hbeta flux to the \oiii $\lambda5007$ flux is $F_\mathrm{\hbeta}/F_\mathrm{[O~III]} = 0.099 \pm 0.002$, which is in good agreement with the value of $F_\mathrm{\hbeta}/F_\mathrm{[O~III]} = 0.110 \pm 0.010$ found by \citet{peterson2004}.

The red spectrum in Figure~\ref{fig:mean_rms_full} shows the rms of the MDM spectra after subtracting the AGN continuum and stellar continuum models from individual spectra so that only the emission-line components remain. This rms spectrum is expected to be a more accurate representation of the emission-line variability than the rms of the full spectra \citep{barth2015}. We show the rms here and not the excess rms defined by Equation~\ref{eq:excessvariance} because, for parts of the spectra dominated by continuum emission, the continuum-subtracted flux could be lower than the total flux uncertainties and the e-rms would be undefined. Thus, we use the excess rms only for the full spectrum and not for individual fit components.

%--------FIGURE--------%
\begin{figure}
\centering
\includegraphics[angle=0,scale=0.62,trim={0.4cm 0.6cm 0.7cm 0.8cm},clip=true]{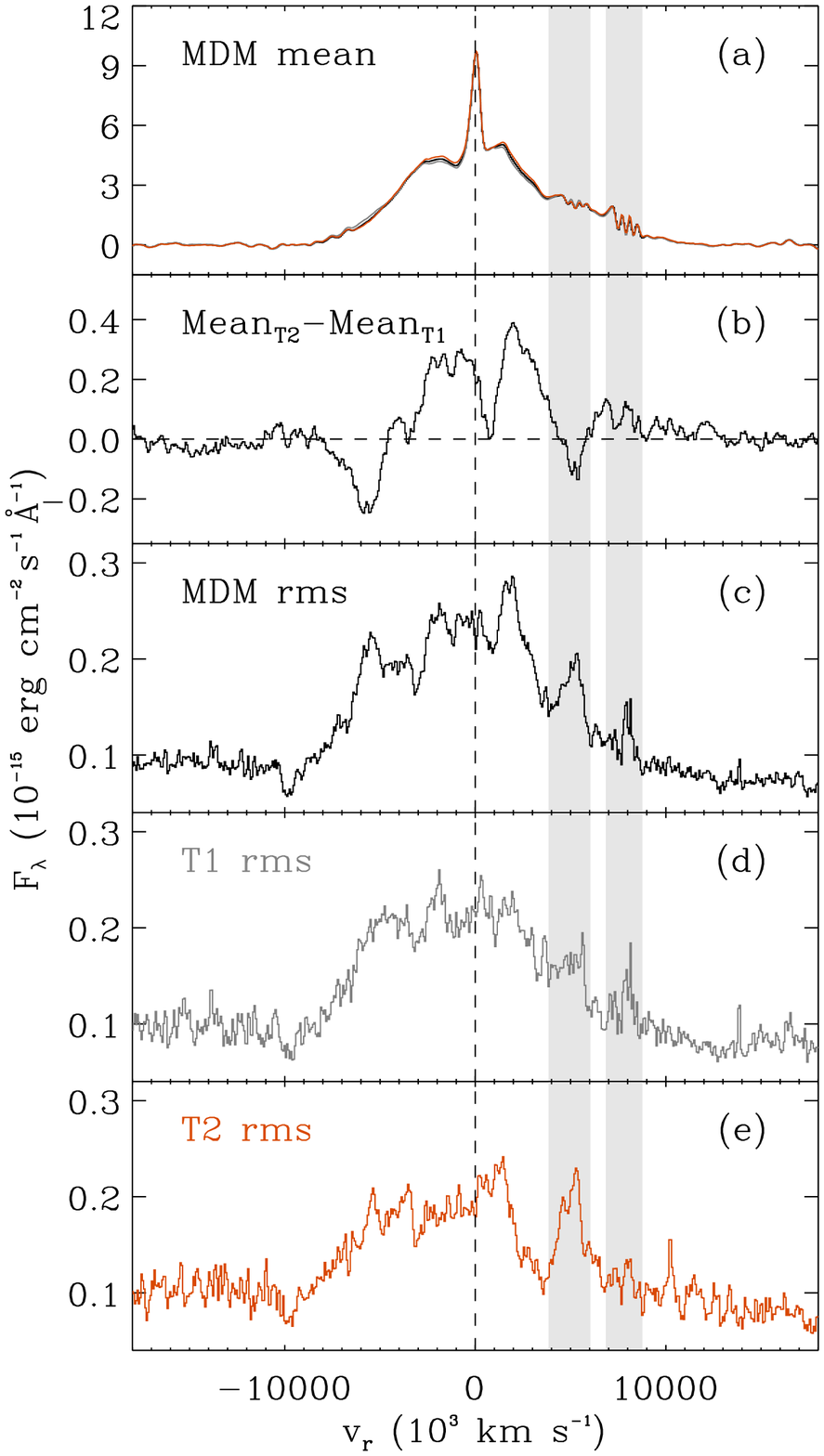}
\caption{The MDM mean spectrum (a), difference between the T1 (THJD $<$ 6747) and T2 (THJD $>$ 6747) mean spectra (b), and the rms spectra (c$-$e) for \hbeta after subtracting all other fit components from spectral decomposition. The colors are for the full campaign (black), T1 (gray), and T2 (orange, see Section 3.3). Zero velocity is determined by the peak of the narrow \hbeta line in the mean spectrum, and the gray bands indicate regions contaminated by \oiii residuals. The rms spectra have statistical uncertainties of $\sim 12\%$.}
\label{fig:mean_diff_rms_fullT1T2}
\end{figure}
%----------------------%

Panels (a) and (c) of Figure~\ref{fig:mean_diff_rms_fullT1T2} show the \hbeta mean and rms fluxes as a function of line-of-sight velocity ($v_\mathrm{r}$) after spectral decomposition, and panel (b) shows the difference between the T1 and T2 mean fluxes. The rms flux has a statistical uncertainty of $\sim 12\%$ and the \oiii residuals are much lower compared to the case with no spectral decomposition (Fig.~\ref{fig:mean_rms_full}). The rms profile still has jagged features, which likely reflect real variability across the broad emission line.

\begin{figure*}
\centering
\includegraphics[angle=0,scale=0.8,trim={0cm 0.4cm 0cm 0.5cm},clip=true]{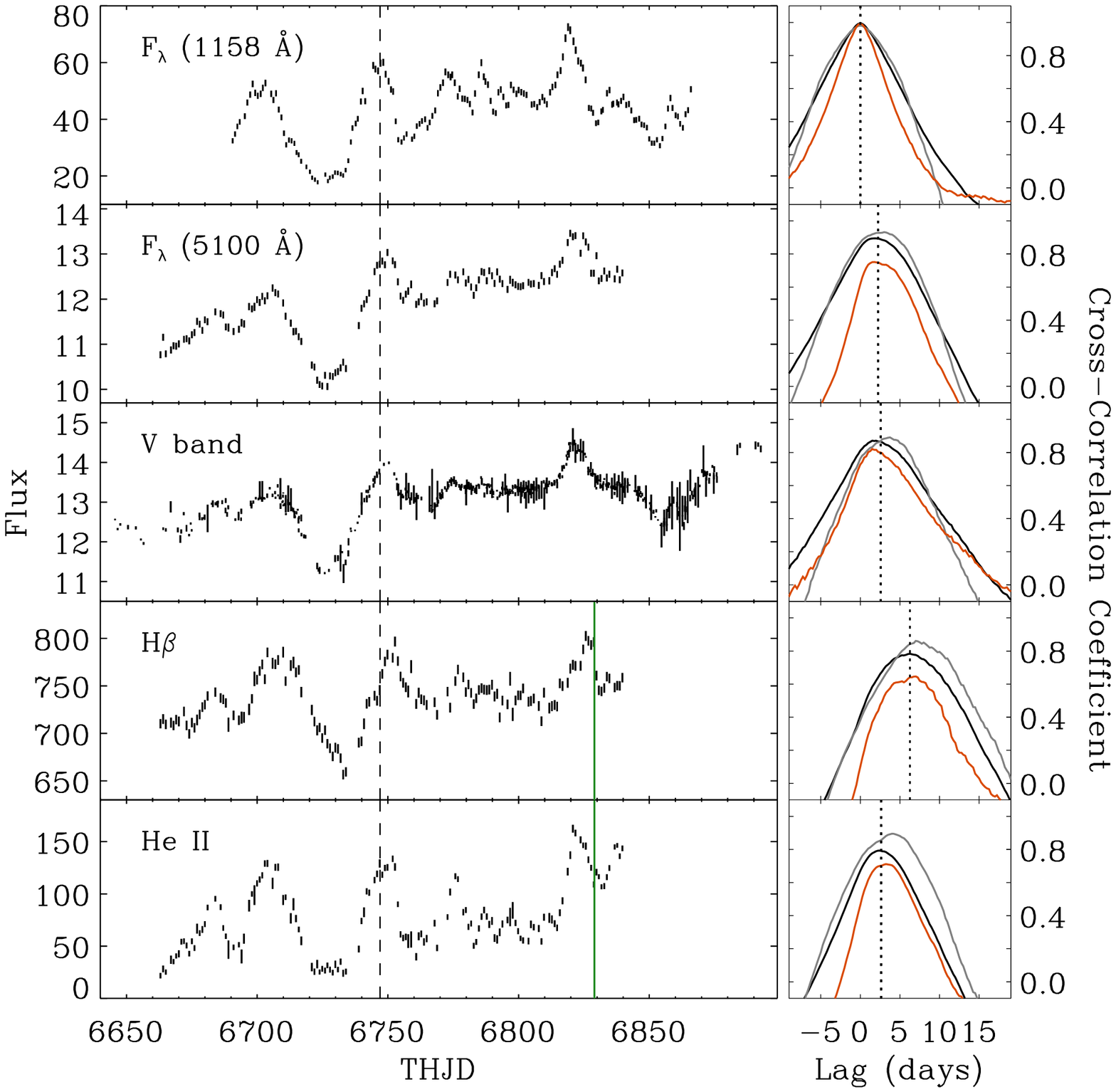}
\caption{\textit{Left:} Light curves for the UV 1158~\angstrom continuum, optical 5100~\angstrom continuum, \vband continuum, \hbeta, and \heiiopt. The continuum light curves are in units of \fluxunitlambda and the line light curves are in units of \fluxunit. The \hbeta and \heii fluxes include contributions from both broad and narrow line components. The solid green vertical line indicates where the emission-line light curves were truncated for the lag analysis (see text) and the dashed black vertical line shows the division between the T1 and T2 periods. \textit{Right:} Cross-correlation functions for each light curve measured against the 1158~\angstrom continuum. The top right panel shows the auto-correlation of the 1158~\angstrom light curve. The black, gray, and orange solid lines represent the CCFs for the full campaign, T1, and T2 (respectively), and the dotted vertical lines denote \taucen for the full campaign.}
\label{fig:lcs_uv_cont_hb_heii}
\end{figure*}

Figure~\ref{fig:lcs_uv_cont_hb_heii} shows the 1158~\angstrom UV continuum light curve from Paper I, the MDM optical 5100~\angstrom continuum light curve, the \vband photometric light curve from Paper III, and the MDM \hbeta and \heiiopt emission-line light curves. The \heii light curve reaches a flat-bottomed minimum near THJD = 6720. This is because the \heii flux includes contributions from the broad- and narrow-line components, so when the broad-line flux is near zero, the total \heii line flux stays at a minimum value equal to the narrow-line flux. Light-curve statistics that quantify the variability of \nagn during the monitoring period are given in Table~\ref{tab:lc_stats}. $F_\mathrm{var}$ is as defined in Equation~\ref{eq:fvar} and $R_{\mathrm{max}}$ is the ratio between the maximum and minimum fluxes.

%--------FIGURE--------%
%----------------------%

\begin{deluxetable*}{lccccc}
  \tablecaption{Statistics for \hst and MDM Light Curves}
  \tablewidth{0pt} 
  \tablecolumns{6}
  \tablehead{
    \colhead{Emission Component} &
    \colhead{Epochs} &
    \colhead{Mean Flux} &
    \colhead{rms Flux} &
    \colhead{$F_\mathrm{var}$} &
    \colhead{\Rmax} 
}
\startdata
    $F_{\lambda}$ (1158 \angstrom)   & 171  &  \phn43.48 $\pm$   0.86  &  11.14 $\pm$ 1.21   & 0.255 & 4.07 $\pm$ 0.18 \\
    $F_{\lambda}$ (5100 \angstrom)   & 133  &  \phn11.96 $\pm$   0.07  & \phn0.80 $\pm$  0.09  & 0.066 & 1.33 $\pm$ 0.01 \\
    \hbeta                        &   133  &    738.49  $\pm$   2.40  & 28.29 $\pm$ 3.38  & 0.038 & 1.22 $\pm$ 0.01 \\
    \heiiopt                      &   133  & \phn78.71  $\pm$   2.95  & 35.14 $\pm$ 4.17  & 0.444 & 7.48 $\pm$ 0.91 \\
\\
    $F_{\lambda}$ (1158 \angstrom, T1) &\phn51  & \phn35.85  $\pm$ 1.79  &  12.61 $\pm$ 2.54  & 0.351  & 3.31 $\pm$ 0.15 \\
    $F_{\lambda}$ (1158 \angstrom, T2) & 120    & \phn46.72  $\pm$ 0.80 & \phn8.66 $\pm$ 1.13  & 0.184  & 2.36 $\pm$ 0.08 \\
    $F_{\lambda}$ (5100 \angstrom, T1) & \phn67 & \phn11.31 $\pm$ 0.08 & \phn0.67 $\pm$ 0.12  & 0.059  & 1.27 $\pm$ 0.01 \\
    $F_{\lambda}$ (5100 \angstrom, T2) & \phn67 & \phn12.51 $\pm$ 0.04 & \phn0.37 $\pm$ 0.06  & 0.029  & 1.13 $\pm$ 0.01 \\
    \hbeta (T1)                       & \phn67 &   725.10  $\pm$ 3.80 & 30.47  $\pm$ 5.36 & 0.041  & 1.20 $\pm$ 0.01 \\
    \hbeta (T2)                       & \phn67 &   750.14  $\pm$ 2.38 & 20.11  $\pm$ 3.33 & 0.026  & 1.13 $\pm$ 0.01 \\
    \heiiopt (T1)                     & \phn67 & \phn65.84  $\pm$ 4.16 & 33.61 $\pm$ 5.89 & 0.507  & 5.95 $\pm$ 0.73 \\
    \heiiopt (T2)                     & \phn67 & \phn89.90  $\pm$ 3.75 & 32.71 $\pm$ 5.34 & 0.362  & 4.07 $\pm$ 0.37 

 \enddata

\tablecomments{Continuum flux densities are in units of \fluxunitlambda and emission-line fluxes are in units of \fluxunit. T1 and T2 denote the first and second halves of the campaign (respectively) divided at THJD = 6747.}
\label{tab:lc_stats}
\end{deluxetable*}

\subsection{Host-Galaxy Flux Removal}

We measured the host-galaxy contribution to the continuum using an ``AGN-free'' image of \nagn generated by \citet{bentz2013} after performing two-dimensional surface brightness decomposition on \textit{HST} images of the galaxy. We found that the amount of starlight expected through a 5\arcsec $\times$ 15\arcsec~aperture with a slit position angle of $0\degree$ is $F_\mathrm{5100, gal} = (4.52~\pm~0.45) \times \fluxunitlambda$. Subtracting this from the mean continuum flux density of $F_\mathrm{5100} = (11.96~\pm~0.07) \times \fluxunitlambda$ gives a mean AGN flux of $F_\mathrm{5100,AGN} = (7.44~\pm~0.50) \times \fluxunitlambda$, which is consistent with the value of $F_\mathrm{5100,AGN} = (7.82~\pm~0.02) \times \fluxunitlambda$ measured from the power-law component of the spectral decomposition for the mean MDM spectrum.

\subsection{Anomalous Emission-Line Light-Curve Behavior}

RM analyses typically assume that the emission-line light curve responds linearly to continuum variations and is a lagged, scaled, and smoothed version of the continuum light curve. However, this does not appear to be the case for a portion of our campaign. As described in Paper I and Paper IV, there are significant differences between the UV continuum and emission-line light curves after THJD = 6780. The continuum flux increased while the emission-line fluxes either decreased or remained roughly constant in a suppressed state for the remainder of the campaign.

\begin{figure}
\centering
\includegraphics[angle=0,scale=0.60,trim={0.0cm 0.4cm 0.0cm 0.0cm},clip=true]{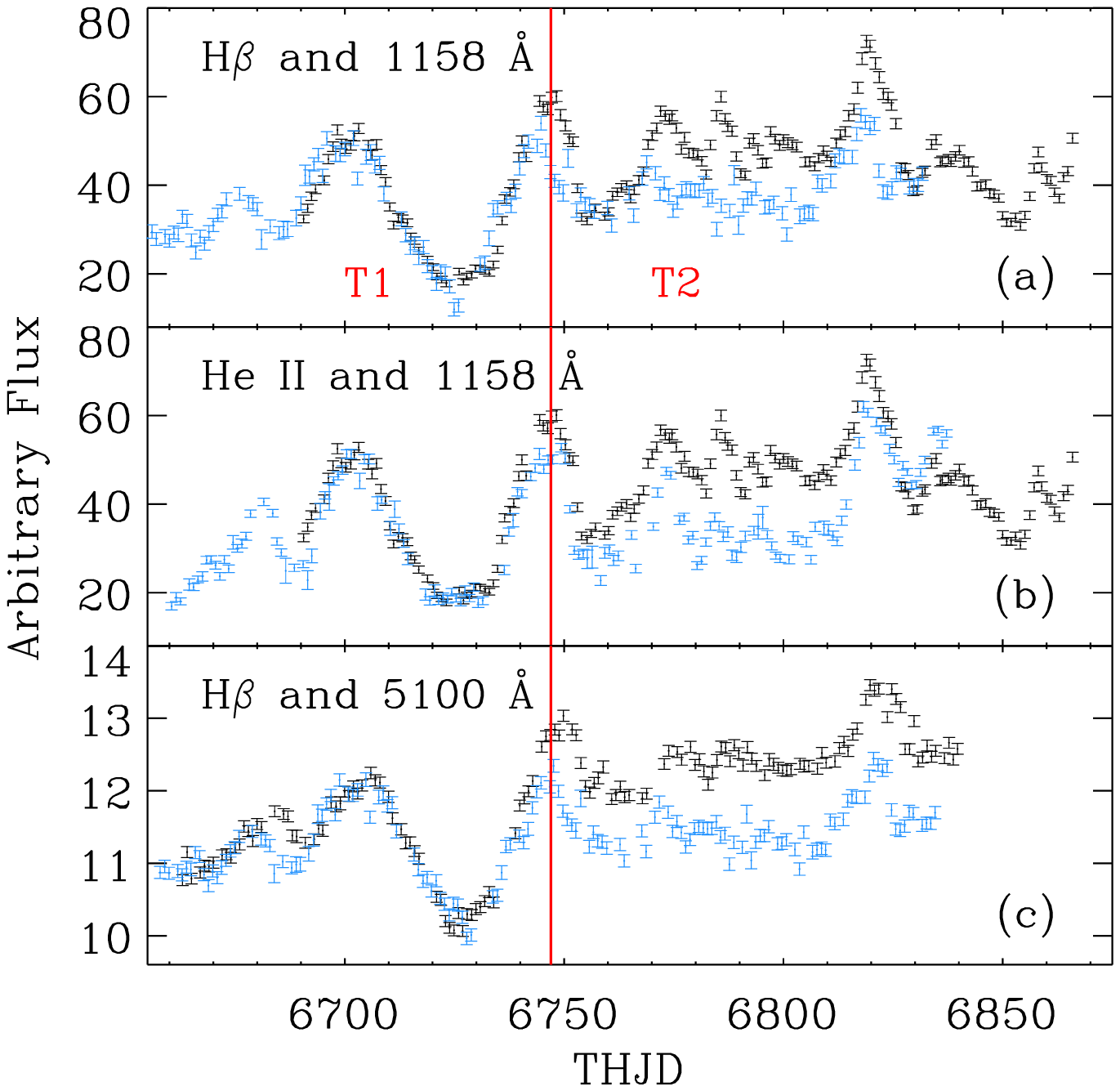}
\caption{1158~\angstrom and 5100~\angstrom continuum light curves (black) compared with scaled and shifted emission-line light curves (blue). The vertical line at THJD = 6747 indicates the epoch separating the T1 and T2 segments. In each of the panels, the emission-line light curve closely tracks the continuum light curve in T1, but appears to correlate less closely with the continuum variations in T2.}
\label{fig:lcs_t1t2}
\end{figure}

The \hbeta emission line shows similar behavior to that of \lya and \civ, in that there is a marked difference in the line response between the first and second halves of the campaign. This ``decorrelation'' phenomenon is illustrated in Figure~\ref{fig:lcs_t1t2}(a), where the \hbeta and UV continuum light curves trace each other well in the first half of the campaign, but the continuum flux continues to trend upward beyond THJD = 6740, while the \hbeta flux begins to fall. For the remainder of the campaign, the \hbeta flux remains in a suppressed state and the light curve does not follow the continuum light curve well in that prominent features in the continuum light curves (e.g. THJD $\approx$ 6770, 6785) are not present in the emission-line light curve. The \heiiopt light curve behaves similarly and decorrelates from the continuum at around THJD = 6760 (Figure~\ref{fig:lcs_t1t2}b). This phenomenon is also apparent when comparing the \hbeta light curve to the 5100~\angstrom continuum, as shown in Figure~\ref{fig:lcs_t1t2}(c). Though there are small differences between the light curves in T1 (such as around THJD 6685), the overall correlation in T1 is significantly better than in T2.

%--------FIGURE--------%
%----------------------%

Owing to this change in the emission-line response, we followed the procedures presented in Paper I for determining the UV emission-line lags, and divided the 5100~\angstrom continuum and optical emission-line light curves into two subsets, labeled T1 and T2, to examine the lag of each segment separately. The subsets are divided at THJD = 6747 and each has 67 epochs. The 1158~\angstrom and \vband continuum light curves were also separated into two segments at THJD = 6747. Note that this is a different dividing epoch from the one used in Paper I. The characteristics of the half-campaign light curves are given in the bottom portion of Table~\ref{tab:lc_stats}. Figure~\ref{fig:mean_diff_rms_fullT1T2} shows the MDM mean and rms spectra for T1 and T2 in gray and orange, respectively. While the three mean spectra look almost identical, the T1 and T2 rms spectra are significantly different, which indicates changes in the amount of variability between the two campaign halves.

%====================================================================================================

\section{Data Analysis}

In the following sections, we examine properties of the BLR by measuring the emission-line responses to continuum variations. We also discuss the anomalous behavior of the emission-line light curves observed during this campaign and BH mass measurements using this dataset. 

\subsection{Emission-Line Lags}

We measured the \hbeta and \heiiopt lags relative to both the 5100~\angstrom continuum and the 1158~\angstrom continuum. All light curves were detrended by subtracting a linear least-squares fit to the data to remove long-term trends that may bias lag calculations \citep{welsh1999}. In this case, we found very weak trends for all the light curves, and detrending has a very small effect ($\sim0.01$ days) on the measured lags. We computed the cross-correlation coefficient $r$ for lags between $-20$ and 40 days in increments of 0.25 days using the interpolated cross-correlation function \citep[ICCF;][]{whitepeterson1994}. Two lag estimates were made for each light-curve pair---the value corresponding to $r_\mathrm{max}$ ($\tau_\mathrm{peak}$) and the centroid of all values with $r > 0.8r_\mathrm{max}$ ($\tau_\mathrm{cen}$). Estimates for the final $\tau_\mathrm{peak}$ and $\tau_\mathrm{cen}$ values and their uncertainties were obtained using Monte Carlo bootstrapping analysis \citep{peterson2004}, where many realizations of the continuum and emission-line light curves were created by randomly choosing $n$ data points with replacement from the observed light curves, where $n$ is the total number of points in the dataset. If a data point is picked $m$ times, then the uncertainty on that point is decreased by a factor of $m^{1/2}$. Each value is then varied by a random Gaussian deviate scaled by the measured flux uncertainty. We constructed $10^3$ realizations of each light curve and computed the cross-correlation function (CCF) for each pair of line and continuum light-curve realizations to create a distribution of \taupeak and \taucen values. The median value from each distribution and the central 68\% interval are then taken to be the final lag and its uncertainty.

\begin{figure}
\begin{center}
\includegraphics[angle=0,scale=0.5,trim={0.2cm 0cm 0cm 0cm},clip=true]{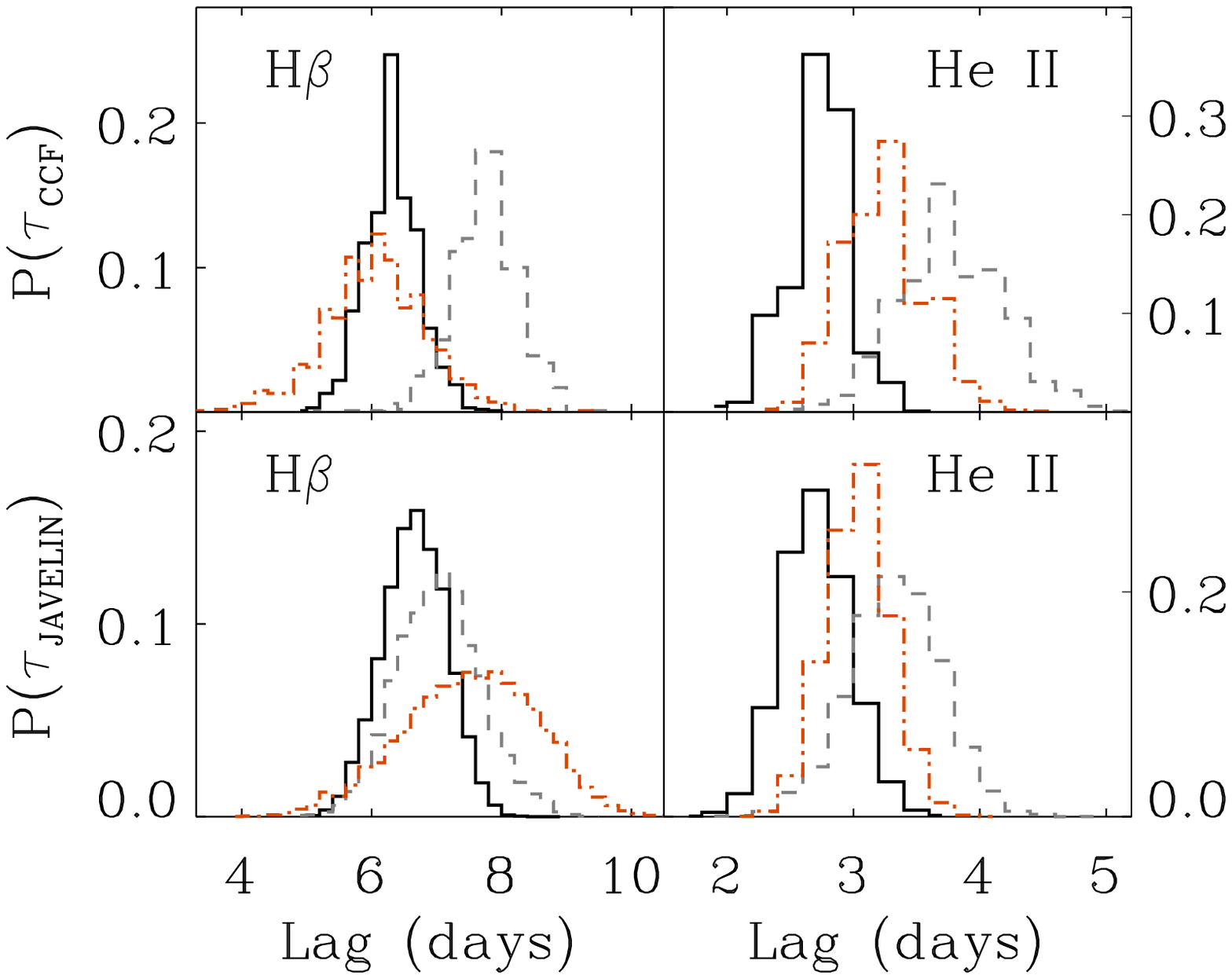}
\caption{\taucen (\textit{top}) and \javelin (\textit{bottom}) lag probability distributions for \hbeta (\textit{left}) and \heii (\textit{right}) measured against the 1158~\angstrom continuum. Black solid lines are for the full campaign, gray dashed lines are for T1, and orange dot-dash lines are for T2.}
\label{fig:ccf_javelin_2x2_HST}
\end{center}
\end{figure}

Table~\ref{tab:lagtable} lists the ICCF lags for \hbeta and \heiiopt measured against the 1158~\angstrom, 5100~\angstrom, and \vband continua. The lag between the 5100~\angstrom and 1158~\angstrom continua is also given. For comparison with Paper I, which presents the UV emission-line lags against the 1367~\angstrom continuum, we also include \hbeta and \heii lags measured against this continuum. Distributions of \taucen values from the Monte Carlo bootstrap analysis using the 1158~\angstrom continuum are shown in the top panels of Figure~\ref{fig:ccf_javelin_2x2_HST}.

%-----TABLE-----%
%\begin{landscape}
\begin{deluxetable*}{lcccc|ccc}
\tablecaption{Rest-Frame Emission-Line Lags}
\tablecolumns{8}
\tablewidth{0pt}
\tablehead{
\colhead{Light Curves} & 
\colhead{\taupeak} & 
\colhead{\taucen} & 
\colhead{\taucenTone} & 
\colhead{\taucenTtwo} &
\colhead{\taujav} &
\colhead{\taujavTone} &
\colhead{\taujavTtwo}
}
\startdata

\hbeta vs. $F_{\lambda}$(1158 \angstrom) & $6.14^{+0.74}_{-0.98}$ & $6.23^{+0.39}_{-0.44}$ & $7.62^{+0.49}_{-0.49}$ & $5.99^{+0.71}_{-0.75}$ & $6.56^{+0.48}_{-0.49}$ & $6.91^{+0.64}_{-0.63}$ & $7.42^{+0.97}_{-1.07}$ \\
\hbeta vs. $F_{\lambda}$(1367 \angstrom) & $5.90^{+0.25}_{-0.74}$ & $5.89^{+0.37}_{-0.37}$ & $7.24^{+0.49}_{-0.48}$ & $5.99^{+0.76}_{-0.82}$ & $6.12^{+0.46}_{-0.47}$ & $6.52^{+0.60}_{-0.57}$ & $7.11^{+1.03}_{-1.06}$ \\
\hbeta vs. $F_{\lambda}$(5100 \angstrom) & $4.42^{+0.98}_{-0.25}$ & $4.17^{+0.36}_{-0.36}$ & $4.99^{+0.40}_{-0.47}$ & $3.10^{+0.77}_{-0.80}$ & $3.84^{+0.57}_{-0.59}$ & $5.15^{+0.68}_{-0.69}$ & $4.78^{+1.13}_{-1.17}$ \\
\hbeta vs. \textit{V} band               & $3.93^{+0.98}_{-0.98}$ & $3.79^{+0.37}_{-0.34}$ & $3.82^{+0.57}_{-0.47}$ & $4.13^{+0.55}_{-0.58}$ & $3.54^{+0.45}_{-0.46}$ & $4.89^{+0.66}_{-0.71}$ & $4.05^{+0.93}_{-0.78}$ \\
 & & & & & & & \\                        
\heii vs. $F_{\lambda}$(1158 \angstrom)  & $2.46^{+0.49}_{-0.25}$ & $2.69^{+0.24}_{-0.25}$ & $3.71^{+0.39}_{-0.38}$ & $3.19^{+0.36}_{-0.35}$ & $2.65^{+0.27}_{-0.27}$ & $3.27^{+0.35}_{-0.35}$ & $2.99^{+0.25}_{-0.26}$ \\ 
\heii vs. $F_{\lambda}$(1367 \angstrom)  & $2.21^{+0.25}_{-0.25}$ & $2.45^{+0.25}_{-0.24}$ & $3.43^{+0.36}_{-0.43}$ & $3.16^{+0.29}_{-0.33}$ & $2.41^{+0.25}_{-0.26}$ & $3.04^{+0.35}_{-0.36}$ & $2.79^{+0.25}_{-0.25}$ \\ 
\heii vs. $F_{\lambda}$(5100 \angstrom)  & $0.49^{+0.25}_{-0.25}$ & $0.79^{+0.35}_{-0.34}$ & $1.21^{+0.28}_{-0.36}$ & $0.85^{+0.36}_{-0.36}$ & $0.16^{+0.37}_{-0.37}$ & $1.13^{+0.51}_{-0.48}$ & $0.85^{+0.38}_{-0.35}$ \\
\heii vs. \textit{V} band                & $0.49^{+0.25}_{-0.49}$ & $0.50^{+0.34}_{-0.26}$ & $0.40^{+0.43}_{-0.39}$ & $1.46^{+0.34}_{-0.27}$ & $0.44^{+0.23}_{-0.23}$ & $0.63^{+0.37}_{-0.36}$ & $0.91^{+0.23}_{-0.21}$ \\
 & & & & & & & \\
$F_{\lambda}$(5100 \angstrom) vs. $F_{\lambda}$(1158 \angstrom)     & $1.97^{+0.25}_{-0.49}$ & $2.23^{+0.31}_{-0.26}$ & $2.55^{+0.28}_{-0.33}$ & $2.77^{+0.41}_{-0.45}$  & & & \\  
 & & & & & & & \\
\hbeta$_\mathrm{full,~MDM}$ vs. $F_{\lambda}$(1158 \angstrom)       & $5.90^{+0.74}_{-0.49}$ & $6.26^{+0.38}_{-0.37}$    & & & & & \\                                              
\hbeta$_\mathrm{full,~all~sites}$ vs. $F_{\lambda}$(1158 \angstrom) & $5.65^{+0.74}_{-0.25}$ & $6.49^{+0.37}_{-0.37}$    & & & & &                                                 

\enddata

\tablecomments{Rest-frame \hbeta and \heiiopt lags (days) for the full campaign and for the T1 (THJD $<$ 6747) and T2 (THJD $>$ 6747) subsets, measured using both ICCF and \javelin. The last two lines show the \hbeta lags measured using the light curve derived without spectral decomposition up to THJD = 6828.75.}
\label{tab:lagtable}
\end{deluxetable*}
%\end{landscape}

We also computed \hbeta and \heiiopt lags for time periods T1 and T2, and found the T1 lags to be consistently longer by about 2$\sigma$. For \hbeta, the T1 and T2 lags bracket the full-campaign lag, while for \heiiopt, the full-campaign lag is shorter than both T1 and T2 lags. For comparison, Paper I found the \lya, \siiv, \civ, and \heii \lam1640 lags to be longer for T2 than for T1, which is the opposite of what we find for \hbeta.

To illustrate the effects of spectral decomposition, we also include in Table~\ref{tab:lagtable} the ICCF lags for the \hbeta light curve where the fluxes were measured using the straight-line continuum-subtraction method and without spectral decomposition. We calculated lags for both the MDM-only and the multi-site \hbeta light curves, which were truncated at THJD = 6828.75 to exclude the last 10 epochs of MDM data (see Section 2). The lags measured with and without using spectral decomposition are consistent to within 1$\sigma$.

%--------FIGURE--------%
%----------------------%

In addition to the ICCF method, we computed the emission-line lags using the \javelin suite of Python codes \citep{zu2011}. We used \javelin to linearly detrend the light curves and model the AGN continuum variability as a damped random walk process \citep[DRW;][]{kelly2009,zu2013}. \javelin explicitly models the emission-line light curves as smoothed, scaled, and lagged versions of the continuum light curve. Since the decorrelation of the line and continuum light curves during the latter half of the campaign clearly violates these assumptions, it is of interest to examine the consequences for the \javelin models. For these models, we simultaneously fit the \hbeta and \heii light curves using either the 1158~\angstrom, 5100~\angstrom, or \vband continuum light curve. The AGN STORM light curves are too short to accurately determine the DRW damping timescale, so this value was fixed to $\tau_\mathrm{DRW} \approx 164$ days as derived by \citet{zu2011} from fits to the 13-year light curve of \nagn \citep{agnwatchXVI}. The precise value of $\tau_\mathrm{DRW}$ is not critical for the algorithm to work provided that it is approximately correct.

%--------FIGURE--------%
\begin{figure}
\centering
\includegraphics[angle=0,scale=0.54,trim={0.2cm 0.8cm 0cm 1.0cm},clip=true]{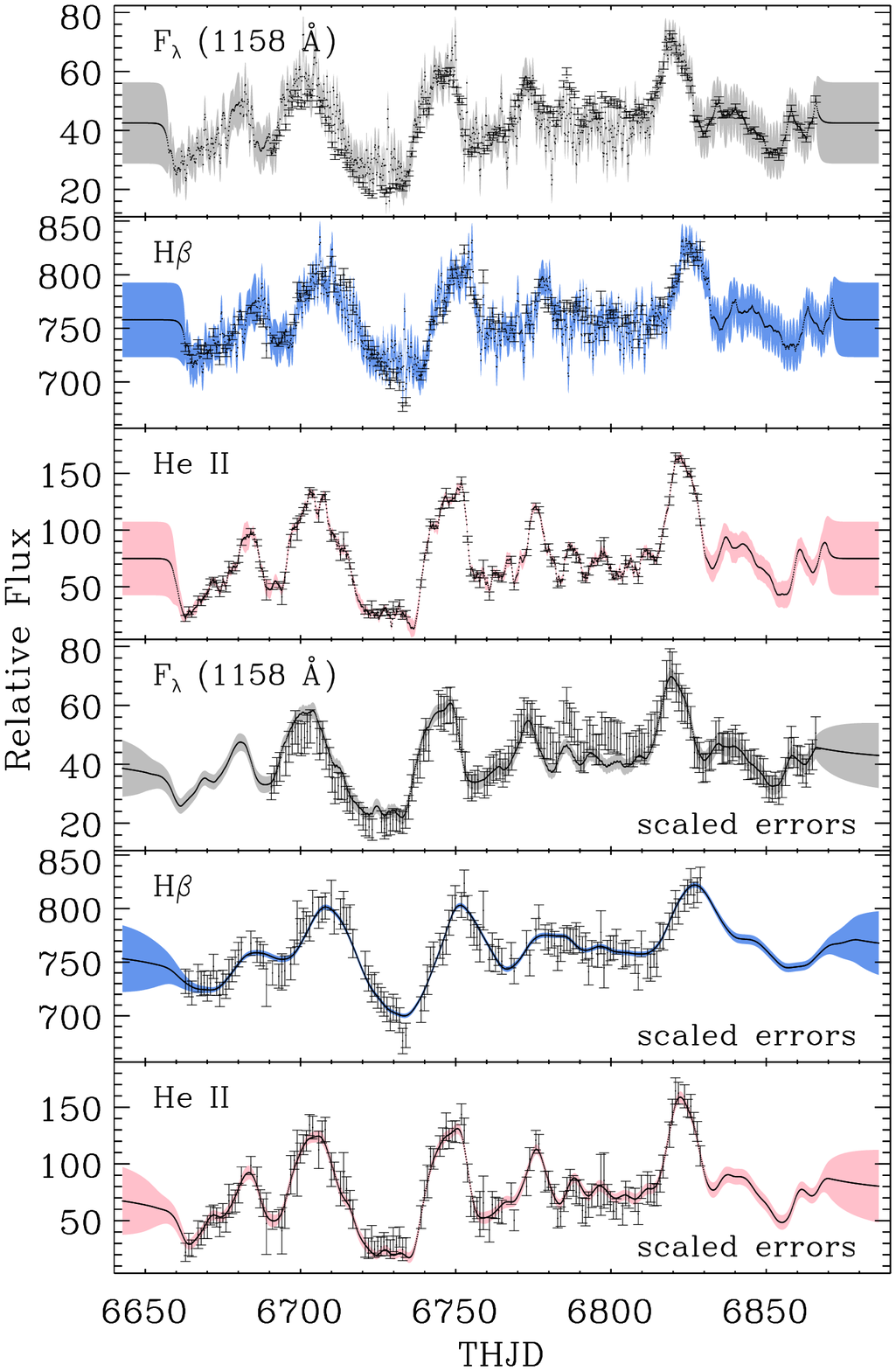}
\caption{\javelin light curves from simultaneously modeling the \hbeta and \heiiopt emission lines with the UV 1158~\angstrom continuum. The data points are measured from observations, the black solid lines are the weighted means of the model light curves consistent with the data, and the thickness of the shaded regions indicates the 1$\sigma$ spread of those light curves. The top three panels show the observed data and \javelin model light curves without any error scaling, and the bottom three panels show the data and light-curve models with scaled uncertainties.}
\label{fig:javelinLCs_allsix}
\end{figure}
%----------------------%

The top three panels of Figure~\ref{fig:javelinLCs_allsix} show the results of using \javelin directly on the observed data. Despite the long DRW time scale, the light-curve models show rapid fluctuations. The algorithm tries to match the suppressed flux in the line light curve to the continuum light curve, and because of the small uncertainties, it strongly prefers lag times for which the continuum and line light curves have minimal temporal overlap. This caused the posterior lag distribution to have multiple narrow peaks corresponding to lags that best desynchronize the light curves.

We can attempt to compensate for this problem by increasing the uncertainties to encompass the amplitude of the decorrelation. This requires scaling up the full-campaign light-curve uncertainties by factors of 5 and 3 for the continuum and line light curves, respectively. The lower panels in Figure~\ref{fig:javelinLCs_allsix} show the \javelin results from fitting to the light curves with scaled errors, where the broader uncertainties allow the algorithm to construct smooth light-curve models. Since there is more statistical weight from fitting both lines simultaneously, the models track the line light curves best and show a smooth systematic offset for the continuum where the line and continuum light curves are decorrelated. When computing the half-campaign lags, \javelin favors a smaller line flux scale factor for T2 to account for the suppressed line fluxes, which begins near the epoch separating T1 and T2. We therefore did not need to scale the flux errors by as much as for the full campaign to account for the decorrelation, and used an error scaling factor of 3 for both continuum and line light curves. The resulting \javelin lags, shown in Table~\ref{tab:lagtable}, are consistent with those measured using the ICCF method. Similar to the ICCF lags, the \javelin lags for the full-campaign light curves are also shorter than those for T1 and T2. However, since the \javelin assumptions of the relationship between the line and continuum light curves are not valid for this campaign and the flux errors were scaled for the sole purpose of producing convergent solutions, we do not use the \javelin lags for subsequent analysis.

%~~~~~~~~~~~~~~~~~~~~~~~~~~~~~~~~~~~~~~~~~~~~~~~~~~~~~~~~~~~~~~~~~~~~~~~~~~~~~~

%--------FIGURE--------%
\begin{figure*}
\centering
\includegraphics[angle=0,scale=0.84,trim={0.0cm 0.6cm 0.0cm 0.4cm},clip=true]{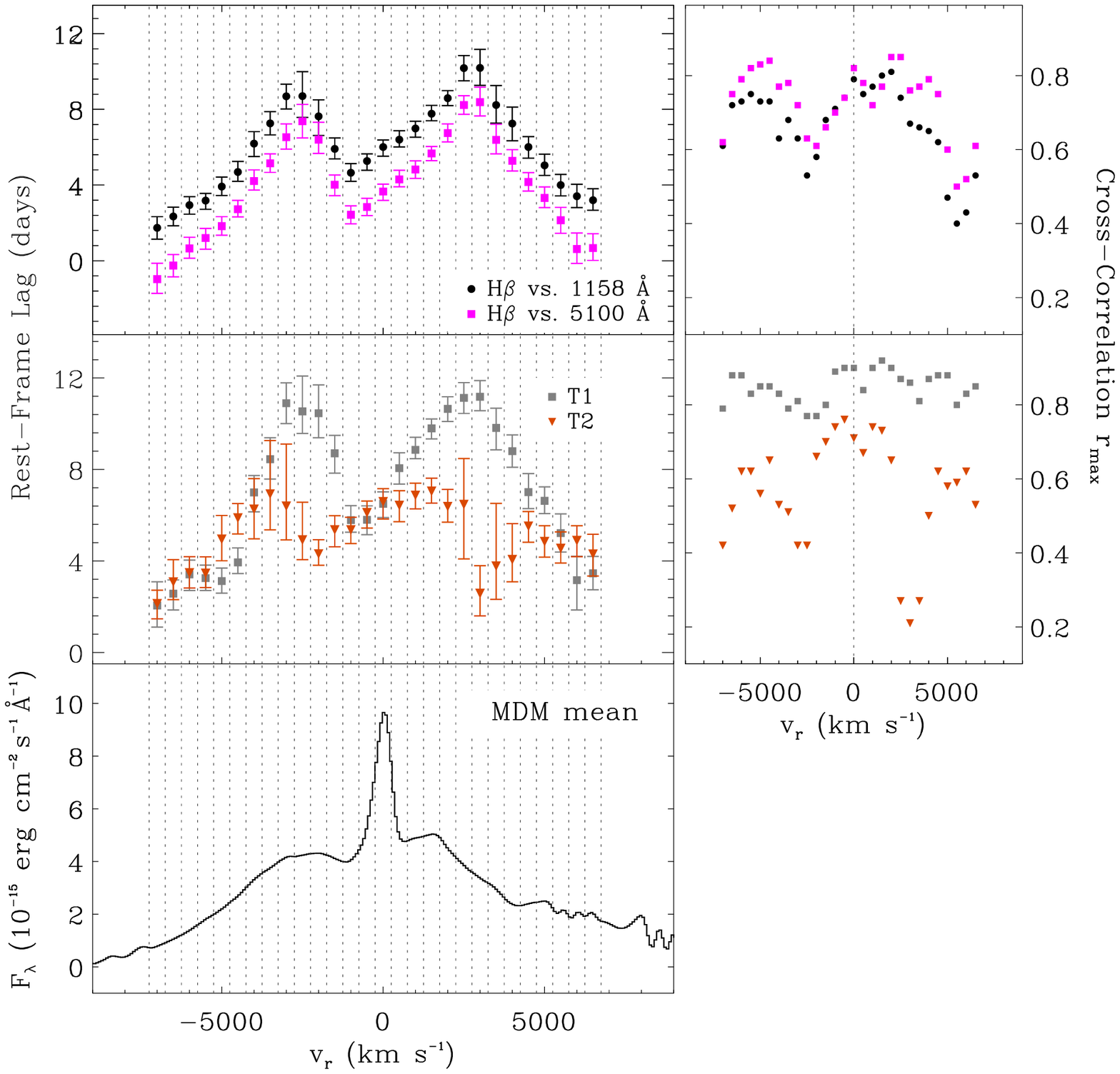}
\caption{\textit{Top left:} ICCF \hbeta lags (\taucen) for 500-\kms bins measured against the 1158~\angstrom and 5100~\angstrom continua. \textit{Middle left:} Lags for T1 and T2 measured against the 1158~\angstrom continuum. \textit{Bottom left:} MDM mean spectrum for the full campaign. \textit{Right:} Maximum cross-correlation coefficients ($r_\mathrm{max}$) for individual velocity bins.}
\label{fig:velres_lags}
\end{figure*}
%----------------------% 

\renewcommand{\thefootnote}{\fnsymbol{footnote}}

\begin{figure}
\centering
\includegraphics[angle=0,scale=0.43,trim={0.15cm 0.3cm 0cm 0.2cm},clip=true]{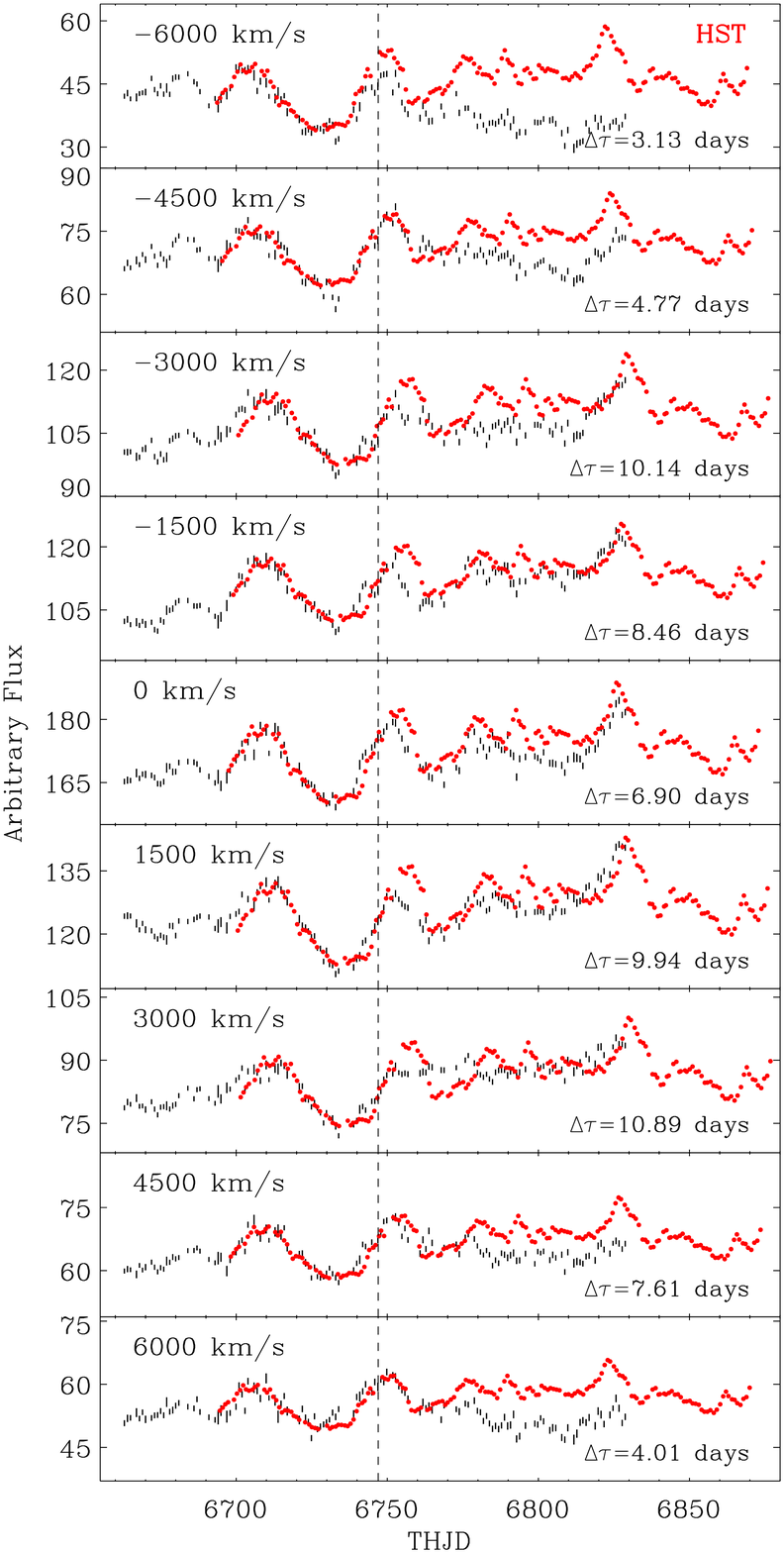}
\caption{Velocity-resolved \hbeta light curves for 1500-\kms bins (black) with the central velocity for each bin shown at the top left of each panel. \textit{HST} 1158~\angstrom light curves that have been scaled and shifted to match the first half of the \hbeta light curves are shown in red, and the dashed line indicates the epoch that separates T1 and T2 for this analysis. The \hbeta response to the continuum is distinctly velocity-dependent during the second half of the campaign.}
\label{fig:lcs_velres_manybins_wHST}
\end{figure}

We examined the velocity-resolved emission-line response by dividing the \hbeta line profile into bins with velocity width of 500~\kms and set zero velocity using the peak of the narrow \hbeta component in the mean spectrum. We constructed light curves for each velocity bin separately, and Figure~\ref{fig:lcs_velres_manybins_wHST} shows the light curves for 1500~\kms velocity bins across the \hbeta line profile (black), with the velocity at the center of each bin shown in the top left of each panel. There were six epochs of spectra\footnote[2]{THJD = [6689.0, 6699.0, 6757.8, 6758.9, 6796.8, 6797.8]} that produced outlying \hbeta fluxes and significantly higher-than-average flux uncertainties for individual velocity bins, so we removed them from the velocity-resolved light curves in order to improve the lag measurements.

We determined the ICCF lag for each of these binned light curves with respect to both the UV and optical continua, and show these lags as a function of line-of-sight velocity in the top left panel of Figure~\ref{fig:velres_lags}. The second left panel shows the velocity-resolved \hbeta--UV lag for T1 and T2, and the maximum cross-correlation coefficients ($r_\mathrm{max}$) are shown in the right-side panels. The bottom panel of Figure~\ref{fig:velres_lags} shows the MDM full-campaign mean spectrum for reference, and Table~\ref{tab:velres_lagtable} lists the velocity-resolved lags. Emission-line variations have lower amplitudes during T2 compared with T1, which led to poorly-constrained ICCF lags with large uncertainties for velocity bins with low line flux. We therefore reduced the upper limit of the CCF lag range from 40 days to 20 days when computing the T2 lags.

The velocity-resolved \hbeta--UV lag for the full campaign is shortest ($\tau_\mathrm{cen} \approx 2$ days) in the line wings where $v_\mathrm{r} \approx \pm7000$~\kms. The lag increases as $v_\mathrm{r}$ approaches zero from both sides of the lag profile, reaches local maxima of $\tau_\mathrm{cen} \approx 10$ days at about $v_\mathrm{r} \approx \pm3000$ \kms, and then steadily decreases until it reaches a local minimum of $\tau_\mathrm{cen} \approx 4$ days near the line profile center. The lag profile measured against the optical continuum has a similar shape, but with all lags $\sim 2$--3 days shorter, as we would expect from the $\sim 2$-day lag between the two continua (Table~\ref{tab:lagtable}). A similar double-peaked lag profile is also observed for \lya (see Paper I). The T1 lag profile closely resembles that of the full campaign, but the T2 lag profile shows a slightly different structure. The bins where the T1 and T2 lags are most discrepant are also where the T2 light curves are least correlated with the continuum ($r_\mathrm{max} < 0.4$). However, even excluding these outliers, there are still discernible differences between the T1 and T2 lag profiles. Additionally, the T2 $r_\mathrm{max}$ values are lower than those of T1 in every velocity bin, which clearly demonstrates that the line and continuum light curves are less correlated in T2 than in T1.

The shape of the velocity-resolved lag profile can provide qualitative information about the kinematics of the line-emitting gas \citep[e.g.,][]{kollatschny2003,bentz2009_lamp,denney2010,barth_zw229,pudu_velres}. In simple models of the BLR \citep{ulrich1984,gaskell1988,welsh1991,horne2004,goad2012,gaskell2013,grier2013}, pure infall motion would lead to longer lags on the blue side of the line profile, and for outflow, the most redshifted gas would have the longest lag. For gas in Keplerian orbits, the shortest lags would be in the line wings, since gas with higher $v_\mathrm{r}$ is closer to the central black hole. Gas with very low $v_\mathrm{r}$ could have a wide range of lags, and a spherical or flat disk distribution of BLR clouds in Keplerian motion could lead to a double-peaked velocity-resolved lag profile if the ionizing source is emitting anisotropically \citep{welsh1991,goadwanders1996,horne2004}.

Previous studies of the UV and optical lines in \nagn have inferred either Keplerian orbits \citep{horne1991,wanders1995,denney2009,bentz2010} or infalling motion \citep{crenshawblackwell1990,donekrolik1996,welsh2007,pancoast2014,gaskellgoosmann2016} for the BLR gas. From our data, the shape of the \hbeta velocity-resolved lag profile suggests a BLR dominated by Keplerian motion. The discrepancy between the T1 and T2 lag profiles may suggest a change in the distribution or dynamics of the BLR gas, though such changes typically occur on timescales much longer than our campaign. More detailed interpretation requires comparison with transfer functions generated for various dynamical models of the BLR, which will be the subject of future work in this series (Pancoast et al., in prep.). 

%--------TABLE--------%
%\begin{landscape}
\begin{deluxetable*}{cccc|cc|cc}
\tablecaption{Rest-Frame \hbeta Velocity-Resolved Lags \label{tab:velres_lagtable}}
\tablecolumns{8}
\tablewidth{0pt}
\tablehead{
\colhead{Wavelength} & 
\colhead{$v_\mathrm{r}$} &
\colhead{\ensuremath{\tau_{\mathrm{Full}}}} & 
\colhead{\ensuremath{r_{\mathrm{max,Full}}}} & 
\colhead{\ensuremath{\tau_{\mathrm{T1}}}} & 
\colhead{\ensuremath{r_{\mathrm{max,T1}}}} & 
\colhead{\ensuremath{\tau_{\mathrm{T2}}}} & 
\colhead{\ensuremath{r_{\mathrm{max,T2}}}} \\
\colhead{(\angstrom)} &
\colhead{(\kms)} &
\colhead{(days)} &
\colhead{} &
\colhead{(days)} &
\colhead{} &
\colhead{(days)} &
\colhead{}
}
\startdata

4743.82$-$4751.92 &    7000      & \phn$1.74^{+0.59}_{-0.60}$ & 0.61 & \phn$2.05^{+1.04}_{-0.94}$ & 0.79 & $2.13^{+0.59}_{-0.66}$ & 0.42 \\
4751.92$-$4760.03 &    6500      & \phn$2.34^{+0.49}_{-0.49}$ & 0.72 & \phn$2.57^{+0.60}_{-0.71}$ & 0.88 & $3.08^{+0.98}_{-0.77}$ & 0.52 \\
4760.03$-$4768.13 &    6000      & \phn$2.94^{+0.45}_{-0.48}$ & 0.73 & \phn$3.41^{+0.63}_{-0.70}$ & 0.88 & $3.48^{+0.71}_{-0.66}$ & 0.62 \\
4768.13$-$4776.23 &    5500      & \phn$3.18^{+0.38}_{-0.46}$ & 0.75 & \phn$3.25^{+0.57}_{-0.54}$ & 0.83 & $3.46^{+0.72}_{-0.62}$ & 0.62 \\
4776.23$-$4784.33 &    5000      & \phn$3.92^{+0.50}_{-0.47}$ & 0.73 & \phn$3.12^{+0.57}_{-0.53}$ & 0.85 & $4.95^{+1.04}_{-0.94}$ & 0.56 \\
4784.33$-$4792.43 &    4500      & \phn$4.69^{+0.56}_{-0.49}$ & 0.73 & \phn$3.94^{+0.63}_{-0.49}$ & 0.85 & $5.88^{+0.63}_{-0.70}$ & 0.65 \\
4792.43$-$4800.53 &    4000      & \phn$6.19^{+0.63}_{-0.68}$ & 0.63 & \phn$7.00^{+0.74}_{-0.85}$ & 0.83 & $6.26^{+1.34}_{-1.29}$ & 0.53 \\
4800.53$-$4808.63 &    3500      & \phn$7.26^{+0.62}_{-0.61}$ & 0.68 & \phn$8.45^{+0.94}_{-0.87}$ & 0.79 & $6.93^{+2.34}_{-1.57}$ & 0.51 \\
4808.63$-$4816.73 &    3000      & \phn$8.69^{+0.64}_{-0.67}$ & 0.63 &    $10.90^{+0.89}_{-0.89}$ & 0.81 & $6.40^{+2.71}_{-1.48}$ & 0.42 \\
4816.73$-$4824.84 &    2500      & \phn$8.70^{+1.29}_{-1.13}$ & 0.53 &    $10.54^{+1.55}_{-0.96}$ & 0.77 & $4.92^{+1.65}_{-0.86}$ & 0.42 \\
4824.84$-$4832.94 &    2000      & \phn$7.63^{+0.87}_{-1.01}$ & 0.58 &    $10.46^{+1.24}_{-1.07}$ & 0.77 & $4.31^{+0.62}_{-0.50}$ & 0.66 \\
4832.94$-$4841.05 &    1500      & \phn$5.91^{+0.58}_{-0.53}$ & 0.68 & \phn$8.71^{+0.78}_{-0.87}$ & 0.80 & $5.36^{+0.62}_{-0.74}$ & 0.70 \\
4841.05$-$4849.15 &    1000      & \phn$4.65^{+0.48}_{-0.45}$ & 0.71 & \phn$5.79^{+0.63}_{-0.53}$ & 0.89 & $5.36^{+0.55}_{-0.60}$ & 0.74 \\
4849.15$-$4857.25 &\phn500       & \phn$5.27^{+0.38}_{-0.49}$ & 0.74 & \phn$5.80^{+0.61}_{-0.65}$ & 0.90 & $6.11^{+0.51}_{-0.67}$ & 0.76 \\
4857.25$-$4865.35 &\phn\phn\phn0 & \phn$6.01^{+0.37}_{-0.49}$ & 0.79 & \phn$6.50^{+0.52}_{-0.60}$ & 0.90 & $6.59^{+0.57}_{-0.76}$ & 0.71 \\
4865.35$-$4873.45 &\phn500       & \phn$6.40^{+0.47}_{-0.39}$ & 0.75 & \phn$8.06^{+0.67}_{-0.70}$ & 0.84 & $6.43^{+0.64}_{-0.71}$ & 0.67 \\
4873.45$-$4881.56 &    1000      & \phn$6.99^{+0.38}_{-0.46}$ & 0.77 & \phn$8.86^{+0.55}_{-0.39}$ & 0.90 & $6.87^{+0.53}_{-0.59}$ & 0.74 \\
4881.56$-$4889.66 &    1500      & \phn$7.77^{+0.44}_{-0.37}$ & 0.80 & \phn$9.80^{+0.41}_{-0.46}$ & 0.92 & $7.05^{+0.57}_{-0.52}$ & 0.73 \\
4889.66$-$4897.76 &    2000      & \phn$8.59^{+0.40}_{-0.39}$ & 0.81 &    $10.65^{+0.53}_{-0.55}$ & 0.90 & $6.38^{+0.75}_{-0.68}$ & 0.65 \\
4897.76$-$4905.86 &    2500      &    $10.18^{+0.66}_{-0.67}$ & 0.74 &    $11.13^{+0.67}_{-0.68}$ & 0.87 & $6.47^{+2.01}_{-2.38}$ & 0.27 \\
4905.86$-$4913.96 &    3000      &    $10.19^{+0.98}_{-0.92}$ & 0.67 &    $11.18^{+0.71}_{-0.61}$ & 0.86 & $2.58^{+1.21}_{-0.98}$ & 0.21 \\
4913.96$-$4922.06 &    3500      & \phn$8.23^{+1.03}_{-0.96}$ & 0.66 & \phn$9.82^{+0.86}_{-0.96}$ & 0.81 & $3.78^{+2.74}_{-1.46}$ & 0.27 \\
4922.06$-$4930.16 &    4000      & \phn$7.25^{+0.87}_{-0.91}$ & 0.65 & \phn$8.80^{+0.71}_{-0.70}$ & 0.87 & $4.06^{+1.57}_{-0.97}$ & 0.50 \\
4930.16$-$4938.27 &    4500      & \phn$6.01^{+0.55}_{-0.59}$ & 0.62 & \phn$7.01^{+0.81}_{-0.72}$ & 0.88 & $5.52^{+0.64}_{-0.70}$ & 0.62 \\
4938.27$-$4946.38 &    5000      & \phn$5.04^{+0.59}_{-0.53}$ & 0.47 & \phn$6.63^{+0.61}_{-0.56}$ & 0.88 & $4.85^{+0.69}_{-0.68}$ & 0.58 \\
4946.38$-$4954.48 &    5500      & \phn$4.00^{+0.56}_{-0.57}$ & 0.40 & \phn$5.22^{+0.84}_{-0.83}$ & 0.80 & $4.55^{+0.72}_{-0.64}$ & 0.59 \\
4954.48$-$4962.58 &    6000      & \phn$3.41^{+0.63}_{-0.60}$ & 0.43 & \phn$3.16^{+1.05}_{-1.30}$ & 0.83 & $4.89^{+0.65}_{-0.70}$ & 0.62 \\
4962.58$-$4970.68 &    6500      & \phn$3.20^{+0.61}_{-0.53}$ & 0.53 & \phn$3.46^{+0.74}_{-0.72}$ & 0.85 & $4.30^{+0.87}_{-0.96}$ & 0.53 

\enddata

\tablecomments{ICCF \hbeta lags (\taucen) for the full campaign and for the T1 (THJD $<$ 6747) and T2 (THJD $>$ 6747) segments. The first column gives the rest-frame wavelength range for each bin and the second column gives the line-of-sight velocity at the center of each bin. The $r_\mathrm{max}$ values give the maxima of the cross-correlation functions between the light curve for each velocity bin and the UV continuum.}
\end{deluxetable*}
%\end{landscape}
%----------------------%

%--------FIGURE--------%
%----------------------%

Figure~\ref{fig:lcs_velres_manybins_wHST} also shows modified versions of the 1158~\angstrom continuum light curve in red. These light curves were shifted in time by the average T1 lag of the bins incorporated in each \hbeta light curve, which are shown in the bottom right of each panel. The fluxes were roughly scaled and shifted to match the first half of the \hbeta light curves. Comparing the line and continuum light curves, it is evident that the \hbeta response in T2 is heavily dependent on the line-of-sight velocity. 

%=============================================================================

\subsection{Anomalous Emission-Line Response to Continuum}

Our data show that previous assumptions about the relationship between the continuum and emission-line light curves---namely that the emission-line light curves are smoothed, scaled, and time-shifted versions of the continuum light curve---are not always valid, as we observed all the UV and optical emission-line light curves decorrelating from the UV continuum about halfway through the monitoring period. Paper IV examined this effect for the UV emission lines by measuring changes in the emission-line equivalent width (EW),

\begin{equation}
\mathrm{EW} = \frac{ F_\mathrm{line}}{F_\mathrm{cont}},
\end{equation}

\noindent and the responsivity (\eff), which is the power-law index that relates the driving continuum flux to the responding emission-line fluxes,

\begin{equation}
\mathrm{log}~F_\mathrm{line} = A + \eff [\mathrm{log}~F_\mathrm{cont}].
\label{eq:eta1}
\end{equation} 

\noindent In the case of no line response, $\eff = 0$, and if the line responds linearly to continuum variations (i.e. the transfer function is a $\delta$ function), then $\eff = 1$. The emission-line EW and the continuum flux are related via the Baldwin relation \citep{baldwin1977}, which is described by

\begin{equation}
\mathrm{log~EW_\mathrm{line}} = B + \beta [\mathrm{log}~F_\mathrm{cont}],
\label{eq:beta1}
\end{equation}

\noindent where the choice for $F_\mathrm{cont}$ is assumed to be a reasonable proxy for the ionizing continuum. Thus, $\beta$ is also known as the slope of the Baldwin relation.

\begin{figure}
\centering
\includegraphics[angle=0,scale=0.59,trim={0.2cm 0.2cm 0.0cm 0.0cm},clip=true]{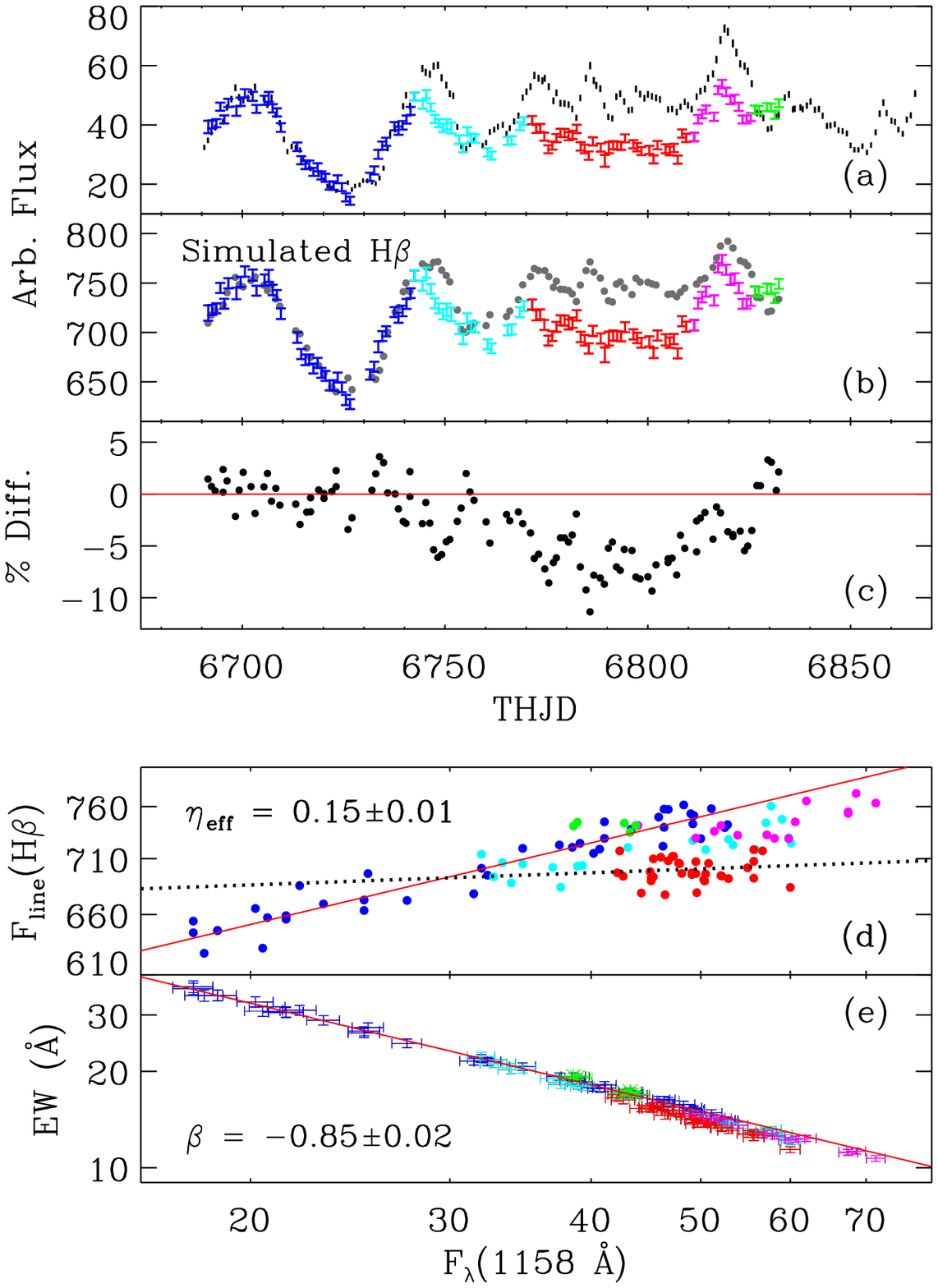}
\caption{Panel (a) shows 1158~\angstrom continuum light curve with the time-shifted \hbeta light curve (see text for color scheme). Panels (b) and (c) show the reconstructed \hbeta light curve (gray dots) and the percent of flux lost during the anomaly, respectively. Panels (d) and (e) show the \hbeta broad-line flux (\fluxunit) and EW (\angstrom) as a function of the 1158~\angstrom continuum flux density (\fluxunitlambda). The solid red lines are linear least-squares fits to the blue points, which are from the period when the line and continuum light curves are well-correlated. The slopes of these fits are shown in each panel. The dotted black line in panel (d) is the least-squares fit to the red points (anomaly) and has a slope of \eff$= 0.02\pm 0.03$.}
\label{fig:holiday_hb_5panel}
\end{figure}

Following the same procedures as in Paper IV, we compute the responsivity \eff and EW for the portion of the \hbeta light curve that correlates with the UV continuum, then examine how these values change in different segments of the light curves. The values $F_\mathrm{cont}$ and $F_\mathrm{line}$ refer to the continuum and emission-line fluxes after removing non-variable components such as host galaxy and narrow-line flux contributions, and after correcting for the mean time delay between the continuum and line light curves \citep{poggepeterson1992,gilbertpeterson2003,goad2004}. There is very little host galaxy flux in the 1158~\angstrom continuum, which is dominated by the variable AGN. For the line fluxes, we took \hbeta fluxes measured after linear continuum removal (without spectral decomposition) and subtracted a constant narrow \hbeta flux measured from the MDM mean spectrum fit to remove the non-variable line component. To correct for the emission-line time delay, we shifted the \hbeta light curve by 8 days, which corresponds to the lag for the portion of the line light curve closely correlated with the UV continuum. Figure~\ref{fig:holiday_hb_5panel}(a) shows the 1158~\angstrom continuum light curve (black) with the time-shifted and flux-scaled \hbeta light curve, which has been truncated at the beginning to match the first epoch of continuum observations. To show the \hbeta light curve's general behavior toward the end of the campaign, we have shown here the full 143 epochs of \hbeta flux measurements instead of the 133-epoch light curve we used in the \hbeta lag analysis. However, since the last 10 epochs of spectra suffer from inconsistent spectral flux calibration (see Section 2), we do not use these points in calculating \eff or $\beta$.

We divided the \hbeta light curve into five segments. The first segment corresponds to the period when the line light curve closely follows the continuum light curve (blue points); the second and third segments correspond to periods when the line light curve decouples from the continuum (cyan points) and remains in a state of depressed flux (red points); the last two segments correspond to the line light curve recovering from the depressed state (magenta points) and correlating once again with the continuum light curve (green points). The epochs that divide these segments are THJD = [6743, 6772, 6812, 6827].
%--------FIGURE--------%
%----------------------%

Figure~\ref{fig:holiday_hb_5panel}(d) and \ref{fig:holiday_hb_5panel}(e) show the \hbeta broad-line flux and EW as a function of the 1158~\angstrom continuum flux density determined from the \hst epoch closest to each MDM epoch. The red lines represent linear least-squares fits to Equations~\ref{eq:eta1} and~\ref{eq:beta1} using only the blue points. We found that the cyan, red, and magenta points---corresponding to when the light curves are not well-correlated---lie well below the best fits of Equations~\ref{eq:eta1} and~\ref{eq:beta1} to the blue points. Furthermore, the epochs during the anomaly (red points) are characterized by an \eff value consistent with zero (\eff$= 0.02 \pm 0.03$, black dotted line in Figure 12d), which shows that the emission-line strength remained constant independent of continuum strength. Similar results were found for the \civ, \lya, \ion{He}{2}(+\ion{O}{3}]) and \ion{Si}{4}(+\ion{O}{4}]) emission lines in Paper IV. Table~\ref{tab:responsivity} summarizes the \eff and $\beta$ values for all broad emission lines measured during this campaign. Note that the values from Paper IV were computed using epochs both before and after the anomaly (blue and green points), whereas our values were calculated using only epochs before the anomaly (blue points).

%--------TABLE--------%
\begin{deluxetable}{cccr}
  \tablecaption{Responsivity, Slope of the Baldwin Relation, and Flux Deficit During the Anomaly for All Emission Lines}
  \tablecolumns{4}
  \tablewidth{0pt}
  \tablehead{
    \colhead{Line ID} &
    \colhead{\eff} &
    \colhead{$\beta$} &
    \colhead{$F_\mathrm{lost}$}
}
\startdata

\lya                      & $0.30 \pm 0.01$ & $-0.73 \pm 0.02$ & 9\% \\
\ion{Si}{4}$+$\ion{O}{4}] & $0.45 \pm 0.01$ & $-0.58 \pm 0.03$ & 23\% \\
\civ                      & $0.25 \pm 0.01$ & $-0.75 \pm 0.01$ & 18\% \\
\ion{He}{2}$+$\ion{O}{3}] & $0.58 \pm 0.04$ & $-0.48 \pm 0.04$ & 21\% \\
\hbeta                    & $0.15 \pm 0.01$ & $-0.85 \pm 0.02$ & 6\% 

\enddata
\tablecomments{All values were measured using the 1158~\angstrom continuum fluxes from this campaign. The first four rows show values from \citet{stormIV}.}
\label{tab:responsivity}
\end{deluxetable}
%----------------------%

We measured the amplitude of the anomaly by comparing the observed \hbeta light curve to a simulated light curve that represents what the line response would have been without the anomaly. The simulated line fluxes ($F_\mathrm{sim}$) were calculated using the observed continuum fluxes and Equation~\ref{eq:eta1}, where \eff and $A$ are chosen to be the best-fit values for the period when the light curves are well-correlated (blue points). This simulated light curve is shown in Figure~\ref{fig:holiday_hb_5panel}(b) with gray dots, and comparing it with the observed data (colored points) shows that the \hbeta line had lower-than-expected variability amplitude and mean flux during T2. If we define the fractional flux loss as $F_{\mathrm{lost}}=(F_{\mathrm{sim}}-F_{\mathrm{obs}}) / F_\mathrm{sim}$ (Figure~\ref{fig:holiday_hb_5panel}c), then this implies an \hbeta flux deficit of $\sim6\%$ during the anomaly (red points), which is close to the deficit for \lya but much smaller than that of the other UV emission lines, as shown in Table~\ref{tab:responsivity}.

\renewcommand{\thefootnote}{\fnsymbol{footnote}}

Table~\ref{tab:goad2004} summarizes the \hbeta responsivities and AGN optical continuum flux densities measured by \citet[][GKK04]{goad2004} for every year of the 13-year monitoring campaign carried out by the AGN Watch consortium \citep{agnwatchXVI}\footnote[3]{\label{note1}The continuum flux densities from \citet{agnwatchXVI} have been updated by \citet{bentz2013} and \citet{kilerci2015}.}, along with the values of $\eff = 0.59 \pm 0.01$ and $\langle F_\mathrm{5100,\mathrm{AGN}}\rangle = (7.44\pm0.50) \times \fluxunitlambda$ calculated from this dataset. Both the responsivity and optical continuum flux density for this campaign are close to those measured from the 1998 data. Using the mean \hbeta flux without narrow-line contributions of $\langle F_\mathrm{\hbeta} \rangle = (690\pm2.40) \times \fluxunit$, we find a mean EW for this campaign of $92.75\pm2.45$~\angstrom, which is lower than values measured by \citet{goadkorista2014} for previous campaigns.

%--------TABLE--------%
\begin{deluxetable}{cccc}
  \tablecaption{\hbeta Responsivity and AGN Optical Continuum Flux Density for \nagn Over 25 Years}
  \tablecolumns{4}
  \tablewidth{0pt}
  \tablehead{
    \colhead{Year} &
    \colhead{\eff} &
    \colhead{Mean $F_\mathrm{cont}$} &
    \colhead{rms ${F_\mathrm{cont}}$}
}
\startdata
1989    & 0.56 $\pm$ 0.04 &  \phn6.54  &   1.27 \\
1990    & 0.84 $\pm$ 0.03 &  \phn3.79  &   0.91 \\
1991    & 0.95 $\pm$ 0.09 &  \phn6.06  &   0.92 \\
1992    & 0.94 $\pm$ 0.05 &  \phn3.34  &   1.17 \\
1993    & 0.43 $\pm$ 0.04 &  \phn5.69  &   0.87 \\
1994    & 0.74 $\pm$ 0.04 &  \phn6.40  &   1.11 \\
1995    & 0.68 $\pm$ 0.04 &  \phn8.71  &   1.01 \\
1996    & 0.54 $\pm$ 0.03 &  \phn7.07  &   1.52 \\
1997    & 0.80 $\pm$ 0.07 &  \phn4.73  &   0.89 \\
1998    & 0.51 $\pm$ 0.02 &    10.05   &   1.44 \\
1999    & 0.41 $\pm$ 0.04 &  \phn8.48  &   1.82 \\
2000    & 0.65 $\pm$ 0.06 &  \phn3.59  &   1.20 \\
2001    & 1.00 $\pm$ 0.12 &  \phn3.65  &   0.86 \\
2014    & 0.59 $\pm$ 0.01 &  \phn7.44  &   0.50 
\enddata
\tablecomments{Flux densities are in units of \fluxunitlambda. The first 13 rows show values from \citet{goad2004}$^\ddagger$, and the last row shows values from this campaign. All \eff values listed were measured with respect to the 5100~\angstrom continuum.}
\label{tab:goad2004}
\end{deluxetable}
%----------------------%

The \hbeta light curve appears to decorrelate from the UV continuum light curve at a somewhat earlier time (THJD $\approx 6742$) than \civ (THJD $\approx 6765$ as found in Paper IV). If we assume that the \hbeta light curve decorrelates and re-correlates with the continuum light curve at the same times as \civ and use the same dividing epochs as those used in Paper IV (THJD = [6766, 6777, 6814, 6830]), then $\eff = 0.13 \pm 0.01$ and $\beta = -0.88 \pm 0.01$ for the blue points.

While \heiiopt also shows anomalous behavior during the T2 period (Fig.~\ref{fig:lcs_t1t2}), its broad line component is very weak and the fitted profile is poorly constrained in the spectral decomposition process. The \heii light curve is thus noisier than that of \hbeta, and the $F_\mathrm{line}$ and EW values are poorly correlated with $F_\mathrm{cont}$. We therefore do not perform a detailed analysis of the responsivity for \heii.

%====================================================================================================

\subsection{Line Width and \mbh Estimate}

The black hole mass in \nagn has been estimated for several previous RM campaigns, including the AGN Watch consortium
\citep{agnwatchXVI}, \citet{bentz2007}, \citet{denney2010}, and \citet{bentz2009_lamp}. The AGN Watch group determined the BH mass using data from each of the program's monitoring years, and subsequent campaigns each produced an independent BH mass value. Here we compute the BH mass using data from this campaign and compare it with previous results.

We measured the line widths from the \hbeta mean and rms spectra as shown in Figure~\ref{fig:mean_diff_rms_fullT1T2}. The narrow \hbeta line essentially disappears in the rms spectrum but is still present in the mean spectrum. We removed this emission component by subtracting a Gaussian fit to the \hbeta narrow line in the mean MDM spectrum, then linearly interpolated over the narrow \hbeta residuals and the \oiii $\lambda$4959 and $\lambda$5007 residuals.

Two emission-line width values are typically measured in RM: the FWHM and the line dispersion, defined by

\begin{equation}
\sigma^{2}_{\mathrm{line}} = \Bigg(\frac{c}{\lambda_0}\Bigg)^2 \Bigg(\frac{\sum \lambda^2_iS_i}{\sum S_i} - \lambda^2_0 \Bigg),
\end{equation}

\noindent where $S_i$ is the flux density at wavelength bin $\lambda_i$ and $\lambda_0$ is the flux-weighted centroid wavelength of the line profile. We measured $\sigma_{\mathrm{line}}$ and FWHM for the mean and rms spectra using the line profile within the rest-frame wavelength range 4669.8$-$5063.0\ignore{obs:4750$-$5150}~\angstrom. We treated the mean profile as double-peaked and followed the procedures described by \citet{peterson2004} to measure its FWHM. From each of the two peaks at 4827.1\ignore{4910}~\angstrom and 4886.1\ignore{4970}~\angstrom, we traced the line profile outward until the flux reached $0.5F_\mathrm{max}$, then traced the line profile inward from the continuum until the flux again reached $0.5F_\mathrm{max}$. The two wavelengths at $0.5F_\mathrm{max}$---which generally agree well for a smooth line profile---are then averaged to obtain the wavelength at half maximum on each side of the profile. The rms profile is more complicated because it has more than two peaks and the troughs between them can reach well below half of the peak fluxes. We therefore identified a single maximum flux $F_\mathrm{max}$ and traced the profile from the continuum toward the center on both sides until the flux reached $0.5F_\mathrm{max}$. The separation between the two wavelengths at $0.5F_\mathrm{max}$ is taken to be the FWHM of the rms spectrum.

The \hbeta line widths and their uncertainties were determined using Monte Carlo bootstrap analysis. With $n$ total spectra, we randomly selected $n$ spectra from the dataset with replacement, constructed mean and rms line profiles, and measured the line dispersion and FWHM. The median and standard deviation of $10^4$ bootstrap realizations are used for the $\sigma_{\mathrm{line}}$ and FWHM and their estimated uncertainties.

There are additional systematic uncertainties in the line widths from using different \feii templates in the spectral decomposition. We repeated the bootstrap analysis after performing spectral decompositions using each of the \citet{bg}, \citet{veron} and \citet{kovacevic} \feii templates, then took the standard deviation of the \hbeta widths from using the different templates as the systematic uncertainty for the line width. This systematic error dominates the error budget for all $\sigma_{\mathrm{line}}$ measurements and the FWHM of the mean spectrum, but is comparable to the uncertainty from bootstrapping analysis for the FWHM of the rms spectrum. We added this systematic uncertainty in quadrature to the statistical uncertainty from the \citet{kovacevic} line width to obtain the final \hbeta line width uncertainty.

The line widths are also affected by instrumental broadening due to the use of a wide slit. The observed line width is the quadratic sum of the intrinsic and instrumental line widths ($\sigma_\mathrm{observed}^2 = \sigma_\mathrm{intrinsic}^2 + \sigma_\mathrm{instrumental}^2$ or similarly for FWHM). To calculate the instrumental broadening for a 5\arcsec~slit, we followed the methods described by \citet{bentz2009_lamp} and compare the \oiii $\lambda$5007 line width measured from our observations to the width measured from a higher-resolution observation taken using a narrow ($\sim2$\arcsec) slit, which represents the intrinsic line width. The FWHM of the \oiii $\lambda$5007 model from the MDM rest-frame mean spectrum is 9.79~\angstrom (572 \kms), while \citet{whittle1992} found a FWHM of 410~\kms using a 2\arcsec~slit. This implies an instrumental broadening of FWHM$_\mathrm{instrument} = $ 399 \kms, corresponding to $\sigma_\mathrm{instrument} = $ 170 \kms for a Gaussian model of the \oiii line profile. We subtract this instrumental width in quadrature from the \hbeta line width measurements to obtain the intrinsic \hbeta line widths, which are listed in Table~\ref{tab:linewidths}. The large uncertainties on the FWHM measurements from the rms spectrum are a result of the jagged shape of the rms profile.

%--------TABLE--------%
\begin{deluxetable}{llcc}
  \tablecaption{\hbeta Rest-Frame Line Widths}
  \tablecolumns{4}
  \tablewidth{0pt}
  \tablehead{
    \colhead{Segment} &
    \colhead{Spectrum} &
    \colhead{$\sigma_{\mathrm{line}}$} &
    \colhead{FWHM} \\
    \colhead{} &
    \colhead{} &
    \colhead{(\kms)} &
    \colhead{(\kms)} 
}
\startdata

 Full & rms  &   4278 $\pm$ 671 &    10161 $\pm$ 587 \\
 T1   & rms  &   4155 $\pm$ 513 &    10861 $\pm$ 739  \\
 T2   & rms  &   4856 $\pm$ 731 & \phn9103 $\pm$ 1279 \\
\\                             
 Full & Mean  &  3691 $\pm$ 162 & 9496 $\pm$ 418   \\
 T1   & Mean  &  3983 $\pm$ 150 & 9612 $\pm$ 427   \\
 T2   & Mean  &  3939 $\pm$ 177 & 9380 $\pm$ 158    

\enddata
\label{tab:linewidths}
\end{deluxetable}
%----------------------%

We use the \hbeta line dispersion measured from the rms spectrum as the velocity dispersion $\Delta V$, as is common practice in RM (\citealt{peterson2004}, but also see \citealt{collin2006} and \citealt{mejiarestrepo2016} for comparison of FWHM and sigma as indicators of BLR virial velocity for BH mass estimation), and calculate the virial product, defined as $\mathrm{VP} = c\tau\Delta V^2/G$. The \textit{f} factor in Equation~\ref{eq:rm}, which incorporates the geometry and kinematics of the BLR, is generally unknown for any individual AGN, so a single value $\langle f \rangle$ is often used to represent the average normalization for all AGNs. This value is usually taken to be the scale factor that puts the sample of RM virial products onto the same \msigma relation as nearby inactive galaxies (see \citealt{kormendyho2013} for a discussion of the uncertainties in the \msigma relation), and can vary depending on the sample of AGNs used in the fit as well as the fitting method \citep{onken2004,watson2007,woo2010,graham2011,park2012,grier2013,hokim2015}. We adopt a value of $\langle f \rangle = 4.47$ as calculated by \citet{woo2015}. Since $\langle f \rangle$ is calibrated using \hbeta lags measured against the optical continuum, we used the \hbeta--5100~\angstrom lag to calculate the VP and BH mass.

Table~\ref{tab:mbh} lists the virial products and the inferred BH masses for the full, T1, and T2 segments. Our \mbh uncertainties do not include uncertainties in $\langle f \rangle$ \citep{woo2015} or scatter in the distribution of $f$ for different AGNs \citep[e.g.][]{hokim2015}. The T1 and T2 \mbh estimates are consistent even though the two lags are different by more than 1$\sigma$. Compared to recent measurements, the full-campaign virial product ($1.49^{+0.49}_{-0.49} \times 10^7$~\msun) is entirely consistent with the values of $1.50^{+0.37}_{-0.51} \times 10^7$~\msun and $1.38^{+0.51}_{-0.41} \times 10^7$~\msun obtained by \citet{bentz2009_lamp} and \citet{kxlu2016}, respectively. Our BH mass measurement ($6.66^{+2.17}_{-2.17} \times 10^7$~\msun) is also consistent to 1$\sigma$ with the dynamical mass of $3.24^{+2.26}_{-0.90} \times 10^7$~\msun as determined by \citet[][P14]{pancoast2014}. P14 use the \vband continuum as the ionizing source but do not use the \textit{f}-factor, which compensates for the difference in lag between using the UV and optical continua (see next section). This will lead to an underestimate of \rblr and hence a proportional underestimate of the BH mass (Eq.~\ref{eq:rm}). If we scale the P14 mass by the ratio between the \hbeta--1158~\angstrom and the \hbeta--\vband lags from Table~\ref{tab:lagtable} (\tauuv/$\tau_{\mathrm{\hbeta}-V} = 1.64 \pm 0.12$), then the two \mbh measurements become much more congruent.

%--------TABLE--------%
\begin{deluxetable*}{lcccccc}
  \tablecaption{\hbeta Line Measurements, \mbh Estimates, and Continuum Flux Densities}
  \tablecolumns{7}
  \tablewidth{0pt}
  \tablehead{
    \colhead{Segment} &
    \colhead{$\sigma_{\mathrm{line}}$} &
    \colhead{\tauoptical} &
    \colhead{Virial Product} &
    \colhead{\mbh} &
    \colhead{$F_\mathrm{5100, total}$} &
    \colhead{$F_\mathrm{5100, AGN}$} \\
    \colhead{} &
    \colhead{(\kms)} &
    \colhead{(days)} &
    \colhead{($10^7$~\msun)} &
    \colhead{($10^7$~\msun)} &
    \multicolumn{2}{c}{(\fluxunitlambda)}
}
\startdata
 Full &  4278 $\pm$ 671 & $4.17^{+0.36}_{-0.36}$ & $1.49^{+0.49}_{-0.49}$ & $6.66^{+2.17}_{-2.17}$ &  11.96 $\pm$ 0.07 &  7.44 $\pm$ 0.50  \\
 T1   &  4155 $\pm$ 513 & $4.99^{+0.40}_{-0.47}$ & $1.68^{+0.44}_{-0.45}$ & $7.53^{+1.96}_{-1.99}$ &  11.31 $\pm$ 0.08 &  6.79 $\pm$ 0.46  \\
 T2   &  4856 $\pm$ 731 & $3.10^{+0.77}_{-0.80}$ & $1.43^{+0.55}_{-0.56}$ & $6.38^{+2.49}_{-2.53}$ &  12.51 $\pm$ 0.04 &  7.99 $\pm$ 0.45
\enddata
\tablecomments{\tauoptical is the rest-frame ICCF \taucen value measured against the 5100~\angstrom continuum.}
\label{tab:mbh}
\end{deluxetable*}
%----------------------%
%==========================================================================

\section{Discussion}

\subsection{Implications of UV and Optical \hbeta Lags}

Ground-based RM campaigns have traditionally used the optical continuum light curve---by necessity---to determine emission-line lags, even though the far-UV continuum is a better proxy for the ionizing source. Lags relative to the UV continuum (\tauuv) thus should yield more accurate estimates of the BLR characteristic radius than lags measured relative to the optical continuum. Our \hbeta--UV lag (\tauuv$=6.23^{+0.39}_{-0.44}$ days) is $\sim2$ days longer than the \hbeta--optical lag (\tauoptical $=4.17^{+0.36}_{-0.36}$ days). Given that past measurements of the  \hbeta-optical lag for this object range from $\sim4$ to $\sim25$ days \citep[][and references therein]{bentz2013}, and assuming that \tauoptical is always $\sim2$ days longer than \tauuv, then the BLR characteristic radius estimated from previous campaigns using only optical data is biased low by $10-50\%$. The difference in these \hbeta lags is also consistent with the optical-to-UV continuum lag of $2.24^{+0.24}_{-0.24}$ days found in Paper III.

Since virial estimates of \mbh scale with \rblr, it may seem that this difference between \tauuv and \tauoptical will change the BH mass estimate for \nagn and other reverberation-mapped AGNs. However, the virial product---not the BH mass---is the quantity that is directly affected, and the normalization factor \textit{f} is still needed to scale the virial product to a calibrated BH mass (Eq.~\ref{eq:rm}). If the ratio of $\tauuv/\tauoptical$ is the same for all AGNs, then all RM virial products would be scaled up by a constant value, so simply changing the value of \textit{f} would remove this bias and leave the RM black hole masses unchanged. Even if the lag ratio is not constant for all AGNs, its effect on the BH mass scale is still small because the largest source of \mbh uncertainty in RM is the calibration uncertainty for \textit{f} at $\sim 0.12$ dex \citep{woo2015}. Furthermore, for \nagn, the \hbeta lag measured during this campaign is relatively short compared to its historical values (see Section 5.3). If the lag had been longer, the ratio of $\tauuv/\tauoptical$ would be closer to unity and the bias in the BLR characteristic size and BH mass estimate would be much smaller.

While the discrepancy between \tauuv and \tauoptical may not change \mbh measurements that use the \textit{f}-factor, dynamical models that directly infer \mbh and BLR characteristics \citep[e.g.,][]{caramel2011,liyanrong2013} using only optical continuum data \textit{are} affected since they do not depend on this virial normalization. The BLR characteristic size for each AGN inferred using optical data alone would thus be biased by a factor that depends on the value of $\tauuv/\tauoptical$ for that particular object. Changes in the inferred \rblr could also impact single-epoch \mbh estimates, which rely on the empirical relation between the BLR characteristic radius and the AGN continuum luminosity \citep[$R_\mathrm{BLR} \propto L_\mathrm{AGN}^{\alpha}$ with $\alpha \approx \nicefrac{1}{2}$, e.g.,][]{laor1998,wandel1999,kaspi2000,mclurejarvis2002,kaspi2005,vestergaardpeterson2006,bentz2006,bentz2009,bentz2013}. If $\tauuv/\tauoptical$ correlates with AGN luminosity, then the expected value of $\alpha$ would change, and if this ratio is different for all AGNs but is uncorrelated with any other AGN properties, then this would introduce additional scatter to the scaling relation.

We can examine the expected scaling of $\tauuv/\tauoptical$ with respect to \mbh and \lagn by using simple disk and BLR ionization models. Assuming \tauuv is the sum of \tauoptical and the optical--UV inter-band continuum lag \tauuvopt, then 

\begin{equation}
\frac{\tauoptical}{\tauuv} = 1 - \frac{\tauuvopt}{\tauuv}.
\label{eq:lagratio}
\end{equation}

\noindent For a standard thin disk, the characteristic size scale of the disk region emitting at wavelength $\lambda$ scales as \ignore{Morgan,kochanek+2010}

\begin{equation}
R_\lambda \propto \mbh^{2/3}l^{1/3}\lambda^{4/3},
\end{equation}

\noindent where $l = L/L_\mathrm{Edd}$. For a simple photoionization equilibrium model, the BLR characteristic radius scales as

\begin{equation}
\rblr \propto \mbh^{1/2}l^{1/2}.
\end{equation}

\noindent If the inter-band continuum lags are due to light-travel time across the accretion disk, then 

\begin{equation}
\frac{\tauuvopt}{\tauuv} \propto \mbh^{1/6}l^{-1/6}\lambda^{4/3}.
\end{equation}

\noindent The lag ratio is thus weakly dependent on both \mbh and accretion rate, where even a factor of $10^3$ increase in the black hole mass or accretion rate will change the ratio by only a factor of three. Empirically, \citet{bentz2013} found a low scatter of 0.13 dex around the \rl relation for 41 AGNs over four orders of magnitude in luminosity, which suggests this effect is indeed small for most AGNs. However, it is important to directly examine the potential consequences of this effect by obtaining more simultaneous observations of the UV/optical continua and the broad emission lines for AGNs over a wide range of luminosities.

%~~~~~~~~~~~~~~~~~~~~~~~~~~~~~~~~~~~~~~~~~~~~~~~~~~~~~~~~~~~~~~~~~~~~~~~~~~~~

\subsection{Anomalous Emission-Line Light Curve Behavior}

Despite apparent differences in the \civ and \hbeta decorrelation start times and flux deficits during T2, it is likely that the cause of the anomalous light-curve behavior is the same for the UV and optical emission lines. The line response during the anomaly is also heavily dependent on the line-of-sight velocity for \hbeta (Fig.~\ref{fig:lcs_velres_manybins_wHST}) and the UV emission lines (Goad et al, in prep.). Paper IV suggests two scenarios that could produce the anomaly: (1) a temporary obscuration of the ionizing source from parts of the BLR by a moving veil of gas between the accretion disk and BLR, or (2) a temporary change in spectral energy distribution of the ionizing source. Future papers will investigate in detail the timing and magnitude of the anomaly for all UV and optical lines using the full multi-wavelength dataset from this campaign.

The bottom panel of Figure~\ref{fig:lcs_t1t2} shows that the \hbeta light curve decorrelation was also clearly detected using only the optical data. If our campaign had lasted for only the duration of T2, we still would have been able to measure the \hbeta lags with fairly high precision ($\tauuv = 5.99^{+0.71}_{-0.75}$ days and $\tauoptical = 3.10^{+0.77}_{-0.80}$ days), but the lag signal would be contaminated by other unknown factors and would lead to a somewhat biased estimate of the BLR characteristic radius. Depending on how common this decorrelation behavior is for \hbeta, this effect could contribute to additional scatter in the single-object \rl relations for \nagn and other AGNs, which can account for about half of the observed scatter in the global \rl relation for the entire sample of reverberation-mapped AGNs \citep{kilerci2015}.

The high cadence and long duration of this campaign have both been crucial in detecting this decorrelation phenomenon. \citet{horne2004} found that a campaign duration of at least three times the maximum BLR light-crossing time is needed to recover high-fidelity velocity-delay maps from reverberation mapping data. Given the \hbeta--UV lag of $\sim 6$ days for \nagn during our monitoring period, the BLR characteristic light-crossing time is $\sim 12$ days and the minimum campaign length needed to recover velocity-delay maps would correspond to $\sim 40$ days, which would not have allowed us to see this decorrelation. In order to detect and characterize these anomalous behaviors in the emission-line light curves, RM campaigns must be much longer than the minimum requirement for obtaining velocity-delay maps.

Finally, while there are no previously published results documenting similar emission-line light curve behavior, it is possible that this decorrelation phenomenon was indeed observed in other AGNs in previous RM campaigns, but were not recognized as such because the campaign had relatively low cadence and/or short duration. RM programs designed to study large numbers of sources with lower cadence would also not be able to detect these decorrelations. This further highlights the importance of high-cadence, long-duration and high-SNR multi-wavelength reverberation datasets in order to determine the prevalence of this phenomenon.

%~~~~~~~~~~~~~~~~~~~~~~~~~~~~~~~~~~~~~~~~~~~~~~~~~~~~~~~~~~~~~~~~~~~~~~~~~~~~~~~~~~~~~~~~~~~~~~~~

\subsection{Comparison to Previous Campaigns: the \rl Relation}

\begin{figure}
\centering
\includegraphics[angle=0,scale=0.55,trim={0.0cm 0.1cm 0cm 0.5cm},clip=true]{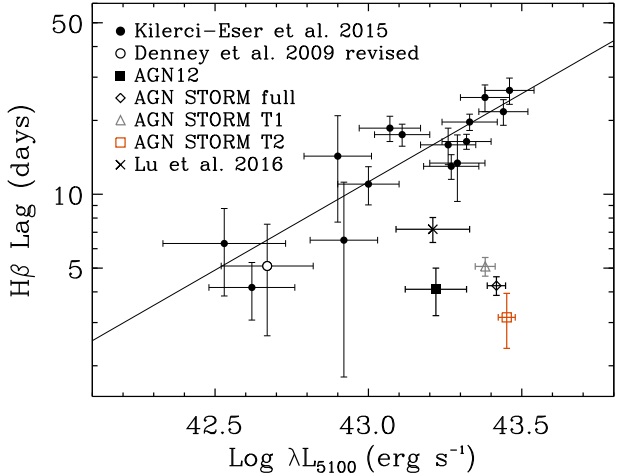}
\caption{\nagn optical continuum luminosity and \hbeta--optical lags from all monitoring campaigns to date. The solid line is the linear least-squares fit to the \citet{kilerci2015} and \citet{denney2009} points.}
\label{fig:rl}
\end{figure}

We compare the \nagn \hbeta--optical lags and optical continuum luminosity (\lfiftyone) from this campaign to those from previous RM campaigns targeting this object, as compiled by \citet{kilerci2015}. The total 5100~\angstrom flux and the AGN continuum flux densities for the full, T1, and T2 segments are listed in columns 6 and 7 of Table~\ref{tab:mbh}, respectively. We applied a Galactic extinction correction of $E(B-V)=0.017$ mag \citep{schlegel1998,schlafly2011} and used a luminosity distance of 75 Mpc in converting fluxes to luminosities. Figure~\ref{fig:rl} shows the \rl relation for \nagn, where the uncertainties are from absolute photometric calibration using \oiii $\lambda5007$ and do not include the luminosity distance uncertainty. The lags from this campaign have smaller uncertainties due to the high cadence and long duration of the ground-based monitoring. \citet{denney2010} monitored \nagn as part of a multi-object RM campaign and found $\tauoptical = 12.40^{+2.74}_{-3.85}$ days. However, the light curves were dominated by a large, long-term trend, and after detrending the light curves with a 3rd order polynomial, the \hbeta lag becomes $5.07^{+2.46}_{-2.37}$~days (open circle in Fig.~\ref{fig:rl}).

%--------FIGURE--------%
%----------------------%

The surprising result is that, given the AGN luminosity during this campaign, the \hbeta lags are nearly five times shorter than expected based on past measurements. \nagn was also monitored in 2012 (AGN12, De Rosa et al., in prep.) and had an \hbeta lag of $4.1 \pm 0.9$~days and a luminosity of $\mathrm{log}(\lambdalfiftyone) = 43.22 \pm 0.1$, whereas \citet{kxlu2016} found the \hbeta lag to be $7.2^{+1.33}_{-0.35}$~days and the AGN luminosity to be $\mathrm{log}(\lambdalfiftyone) = 43.21 \pm 0.1$ during their 2015 campaign, as shown in Figure~\ref{fig:rl}. Both of these lags are also shorter than expected based on past results, though the point from \citet{kxlu2016} suggests that the AGN may be returning to its previously measured \rl relation.

In numerical simulations of the emission-line response to continuum variations for a model BLR with fixed radial extent, \citet{goadkorista2014} found that if the characteristic continuum variability timescale ($\tau_\mathrm{char}$) is smaller than the BLR light-crossing time, then there exists a strong correlation between $\tau_\mathrm{char}$, the line responsivity (\eff), and measured lag, such that both \eff and the lag decrease as $\tau_\mathrm{char}$ decreases (their Figure 9). This is because short timescale continuum variations only probe the inner parts of the BLR, so the measured lag and responsivity would be biased low. If we use the FWHM of the continuum light curve auto-correlation function as a crude proxy for the characteristic variability time scale, then $\tau_\mathrm{char} \sim 10$~days, which is significantly shorter than the values measured for this source for the 1989 \textit{IUE} campaign and the 1993 \hst campaign, and could explain the shorter than expected \hbeta lags.

Regardless of the physical causes of the short lags, these results suggest that the \rl relation is more complex than previously realized. \citet{bentz2013} found a tight correlation between \rblr and \lagn for a sample of $\sim40$ reverberation mapped AGNs, but recent studies by \citet{du2015} and \citet{du2016} have shown that many AGNs with very high accretion rates tend to have considerably shorter \hbeta lags compared to low-accretion AGNs with similar luminosities. Now, we have shown that even for a single AGN with low accretion rate \citep[$\lambda_\mathrm{Edd} = 0.021$ for \nagn;][]{vasudevan2010}, the \rl relation does not always follow a tight power law and that more complex physical processes may contribute significantly to the scatter. In order to further investigate the single-object \rl relation, repeated monitoring campaigns for individual AGNs are needed to track the behavior of each object over a range of timescales and luminosity states. This will, in turn, help improve our understanding of the global \rl relation.

%=====================================================================================

\section{Summary}

We present the results of an optical spectroscopic monitoring program in 2014 targeting the galaxy \nagn as part of the AGN STORM project. Our campaign spanned six months and observed the AGN with an almost daily cadence. Our main findings are as follows.

(1) We determined \hbeta and \heiiopt emission-line lags relative to the far-UV and optical continua, and found that the lag measured against the UV continuum is $\sim2$ days longer than that measured against the optical continuum, consistent with the lag between the UV and optical continua. Given that past measurements of the \hbeta lag against the optical continuum for this object range from $\sim4$ to $\sim25$ days and assuming that this 2-day lag difference is constant over time, then the characteristic size of the BLR inferred from previous data is biased low by 10\%--50\%. Depending on how the ratio of UV and optical \hbeta lags scales with other AGN properties, the RM black hole mass scale and the \rl relation may be affected, which would, in turn, impact single-epoch \mbh estimates for high-redshift AGNs.

(2) We measured velocity-resolved lags for the broad \hbeta line and found a double-peaked lag profile as a function of line-of-sight velocity, with shorter lags in the high-velocity wings. The overall shape of the lag profile is qualitatively similar to those of Keplerian models \citep[e.g.,][]{horne2004}, and is very similar to what is found for \lya \citep{stormI}.

(3) Both the \hbeta and \heiiopt emission lines exhibit significant changes in their response to UV continuum variations halfway through our monitoring campaign. The line light curves decorrelate from that of the continuum and remain in a suppressed state until near the end of the campaign. The same anomalous behavior is observed for all the UV emission lines \citep{stormI,stormIV}. Further investigation into the simultaneous UV and optical line responses during this campaign may elucidate the cause of this anomaly. Depending on how frequently this phenomenon occurs in the AGN population as a whole, this effect could contribute to the scatter in both single-object and global \rl relations. This type of anomalous line behavior is likely only detectable with monitoring campaigns that have a combination of high cadence, long duration, and high data quality.

(4) Given the optical luminosity of \nagn during our campaign, the \hbeta lag measured against the optical continuum is a factor of five shorter than the expected value based on the \rl relation for \nagn from past monitoring campaigns. Our results, combined with other recent \hbeta lag measurements, suggest that this object does not follow a simple power-law \rl relation at all times.

\bigskip

We thank the staffs at the various observatories used to obtain the data in this paper. Support for {\it HST} program number GO-13330 was provided by NASA through a grant from the Space Telescope Science Institute, which is operated by the Association of Universities for Research in Astronomy, Inc., under NASA contract NAS5-26555. L.P. and A.J.B. have been supported by National Science Foundation (NSF) grant AST-1412693. M.M.F., G.D.R., B.M.P., C.J.G., and R.W.P. are grateful for the support of NSF grant AST-1008882 to The Ohio State University. M.C. Bentz gratefully acknowledges support through NSF CAREER grant AST-1253702 to Georgia State University. A.V.F.'s group at UC Berkeley is grateful for financial assistance from NSF grant AST-1211916, the TABASGO Foundation, and the Christopher R. Redlich Fund. C.S.K. is supported by NSF grant AST-1515876. V.N.B. gratefully acknowledges assistance from a NSF Research at Undergraduate Institutions (RUI) grant AST-1312296. M.C. Bottorff acknowledges HHMI for support through an undergraduate science education grant to Southwestern University. K.D.D. is supported by an NSF Fellowship awarded under grant AST-1302093. M.E. thanks the members of the Center for Relativistic Astrophysics at Georgia Tech and the Department of Astronomy at the University of Washington, where he was based during the observing campaign, for their warm hospitality. R.E. gratefully acknowledges support from NASA under awards NNX13AC26G, NNX13AC63G, and NNX13AE99G. J.M.G. gratefully acknowledges support from NASA under award NNH13CH61C. P.B.H. is supported by NSERC. K.H. acknowledges support from the UK Science and Technology Facilities Council through grant ST/M001296/1. TW-SH is supported by the DOE Computational Science Graduate Fellowship, grant number DE-FG02- 97ER25308. M.I. acknowledges support from the Creative Initiative program, No. 2008-0060544, of the National Research Foundation of Korea (NRFK) funded by the Korean government (MSIP). M.D.J. acknowledges NSF grant AST-0618209. SRON is financially supported by NWO, the Netherlands Organization for Scientific Research. B.C.K. is partially supported by the UC Center for Galaxy Evolution. C.S.K. acknowledges the support of NSF grant AST-1009756. D.C.L. acknowledges support from NSF grants AST-1009571 and AST-1210311. P.L. acknowledges support from Fondecyt grant No. 1161184. A.P. acknowledges support from a NSF graduate fellowship and a UCSB Dean’s Fellowship. J.S.S. acknowledges CNPq, National Council for Scientific and Technological Development (Brazil) for partial support and The Ohio State University for warm hospitality. N.T. acknowledges support from CONICYT PAI/82140055. T.T. has been supported by NSF grant AST-1412315. T.T. and B.C.K. acknowledge support from the Packard Foundation in the form of a Packard Research Fellowship to T.T. Also, T.T. thanks the American Academy in Rome and the Observatory of Monteporzio Catone for kind hospitality. The Dark Cosmology Centre is funded by the Danish National Research Foundation. M.V. gratefully acknowledges support from the Danish Council for Independent Research via grant no. DFF 4002-00275. J.-H.W. acknowledges support by the National Research Foundation of Korea (NRF) grant funded by the Korean government (No. 2010-0027910). This work is based partly on observations obtained with the Apache Point Observatory 3.5-m telescope, which is owned and operated by the Astrophysical Research Consortium. Research at Lick Observatory is partially supported by a generous gift from Google. This research has made use of the NASA/IPAC Extragalactic Database (NED), which is operated by the Jet Propulsion Laboratory, California Institute of Technology, under contract with the National Aeronautics and Space Administration. We thank the anonymous referee for the thorough review and helpful comments that helped to improve the clarity of this manuscript.

\bibliographystyle{apj_hack}
\bibliography{citations}

\end{document}